\documentclass[twocolumn]{aastex631}

\usepackage{amsmath, amssymb, xspace, booktabs}

\newcommand{\p}{\prime}
\newcommand{\mrm}{\mathrm}
\newcommand{\ie}{{i.e.}\xspace}

\newcommand{\mbeq}{\overset{!}{=}}

\begin{document}

\title{Subphotospheric Emission from Short Gamma-Ray Bursts. II.~Signatures of Non-Thermal Dissipation in the Multi-Messenger Signals}

\email{mail@annikarudolph.de}

\author[0000-0003-2040-788X]{Annika Rudolph}
\affil{Niels Bohr International Academy and DARK, 
Niels Bohr Institute, University of Copenhagen, 
Blegdamsvej 17, 
2100, Copenhagen, Denmark}

\author[0000-0001-7449-104X]{Irene Tamborra}
\affil{Niels Bohr International Academy and DARK, 
Niels Bohr Institute, University of Copenhagen, 
Blegdamsvej 17,
2100, Copenhagen, Denmark}

\author[0000-0003-3115-2456]{Ore Gottlieb}
\affil{Center for Computational Astrophysics, Flatiron Institute, New York, NY 10010, USA}
\affiliation{Department of Physics and Columbia Astrophysics Laboratory, Columbia University, Pupin Hall, New York, NY 10027, USA}

\begin{abstract}

Building on a general relativistic magnetohydrodynamic simulation of a short gamma-ray burst (sGRB) jet with initial magnetization $\sigma_0=150$, propagating through the dynamical ejecta from a binary neutron star merger, we identify regions of energy dissipation driven by magnetic reconnection and collisionless sub-shocks within different scenarios. We solve the transport equations for photons, electrons, protons, neutrinos, and intermediate particles up to the photosphere,  accounting for all relevant radiative processes, including electron and proton acceleration, and investigate  the potential impact of magnetic reconnection occurring in different regions along the jet. We find the photon spectra  undergo non-thermal modifications below the photosphere, observable in both on-axis and off-axis emission directions, as well as across different  scenarios of energy dissipation and subsequent particle acceleration. Interestingly, the spectral index of the  photon  energy distribution can at most vary by $\sim20\%$ across all different dissipation scenarios. Depending on the dissipation mechanism at play, neutrino signatures may accompany the photon signal, pointing to efficient proton acceleration and shedding light on jet physics. Although our findings are based on one jet simulation, they point to a potential universal origin of the non-thermal features of the Band spectrum observed in sGRBs.
\end{abstract}


\section{Introduction}
\label{sec:intro}
The joint detection of the gravitational wave event GW 170817 with the short gamma-ray burst (sGRB) 170817A~\citep{LIGOScientific:2017zic,LIGOScientific:2017ync,Mooley:2017enz, Goldstein:2017mmi,Savchenko:2017ffs}  provided the first evidence  that binary neutron star mergers can harbor ultra-relativisitc jets~\citep{Eichler:1989ve,Paczynski1991,Nakar:2007yr,Baiotti:2016qnr,Kulkarni:2005jw}. GW 170817 revealed that the interaction between the jet and the ejecta from the merger plays an important role in shaping the jet structure, composition, magnetization and dynamics~\citep{Gottlieb:2020structure,Gottlieb2021:structure,Gottlieb:2022sis,Gottlieb:2023est,Kiuchi:2023obe,Nativi:2021qzr}. While the first detection of a binary neutron star (BNS) merger shed light on the origin and structure of jets, the processes shaping the non-thermal, Band-like spectra observed from GRBs  remain a topic of active debate~\citep[for a recent review see, e.g.][]{Bosnjak:2022sxt}.

In the optically thin regime, synchrotron radiation of accelerated electrons is usually invoked, although  a fine-tuning of the model parameters/physical conditions is required  to  match the observed low-energy spectra and peak energies. Specifically, the fast-cooling photon spectral slope of -1.5 is not compatible with a large number of events, where the marginally fast cooling regime \citep{Daigne:2010fb, Beniamini:2018tbx} or spectral modifications due inverse Compton scatterings in Klein-Nishina regime \citep{Daigne:2010fb} have been proposed to achieve spectral slopes with index $-1$.
A viable alternative is offered by photospheric models, according to which the observed GRB emission is composed of the (modified) thermal spectra released when the jet becomes optically thin. To reproduce the observed  GRB spectra, energy dissipation mechanisms  modifying the spectral energy distribution of photons are typically required~\citep[see][and references therein]{Peer:2016mqn, Beloborodov:2017use}. 

In the optically thick regime (below the photosphere), shocks are expected to be mediated by photons rather than plasma instabilities. Although radiation mediated shocks (RMS) are inefficient at accelerating particles~\citep{Levinson:2019usn},  the superposition of several shocks~\citep{Levinson:2012zy, Keren:2014dsa} or efficient Comptonization in a photon-rich RMS~\citep{Ito:2017jqr, Lundman:2018mad}  may  reproduce the observed low-energy tail of the photon spectra and peak energies~\citep{Samuelsson:2022fbl}. 
Alternatively, inverse Compton and cyclo-synchrotron emission of non-thermal electrons have been identified as additional photon sources to modify their thermal  spectral energy distribution~\citep{Giannios:2007yj, Vurm:2012be, Vurm:2015yfa, Beniamini:2017fqh, Bhattacharya:2019pwc,Gill:2020oon}.
Electrons may be accelerated, e.g.~via magnetic reconnection induced through striped-wind scenarios~\citep{Giannios:2007yj, Beniamini:2017fqh, Gill:2020oon} or collisionless sub-shocks embedded in the  RMS structure~\citep[see e.g.][]{Beloborodov:2016jmz, Ito:2017jqr}. 
Moreover, the shape of the photon spectral energy distribution may also be affected by  neutrons colliding on the protons contained in the flow and inducing a  particle cascade~\citep{Beloborodov:2009be, Vurm:2011fq}. 

Owing to efficient Comptonization, the low-energy tail of the photon spectral energy distribution  does not carry clear signatures of particle acceleration. On the other hand, the high-energy power-law induced by synchrotron emission is directly related to the power-law index of their parent particles. 
These, however, can  only be observed if produced close enough to the photosphere (at optical depth $\tau_T \gtrsim 1$). If the dissipation occurs deeper within the flow, neutrinos, which are  a natural by-product in case of baryon co-acceleration or proton-neutron collisions, could be the only tracers of energy dissipation. 

Considerable effort has been recently devoted to connect magneto-hydrodynamic (MHD) simulations of jets to the modeling of radiation emitted from the jet~\citep{Ito2015,Parsotan2018, Parsotan:2020ilh, Parsotan:2021xki, Ito:2021asl, Ito:2023yhh,Rudolph:2023auv}. 
This can be done by post-processing MHD simulations with a Monte-Carlo technique~\citep{Ito2015,Parsotan2018, Parsotan:2020ilh, Parsotan:2021xki, Ito:2021asl, Ito:2023yhh} or solving the particle and transport equations on top of hydrodynamic and thermodynamic backgrounds  from MHD simulations~\citep{Rudolph:2023auv}. In this context, \cite{Rudolph:2023auv} showed that the subphotospheric photon emission is distorted by non-thermal processes occurring below the photosphere, modifying the otherwise expected Wien photon spectrum. Moreover, 
building   on general-relativistic (GR)MHD simulations of BNS mergers and collapsars,  \cite{Gottlieb:2021pzr, Guarini:2022hry}  showed that neutrinos should be produced with energy $\lesssim 10^5$~GeV because of the large baryon loading of the jet caused by the strong mixing with the shocked medium (``cocoon''), in contrast to simpler analytical models~\citep{Razzaque:2003uv,Ando:2005xi, Murase:2008sp, Wang:2008zm, Gao:2012ay, Asano:2013jea, Murase:2013hh,Tamborra:2015fzv, Xiao:2017blv} which forecasted neutrinos with TeV--PeV energy. 

This paper expands on our earlier {\it Astrophysical Journal Letter}~\citep{Rudolph:2023auv}, where we have solved the coupled transport equations for photons, protons, electrons, neutrinos, as well as intermediate particle species, taking into account all radiative processes and building on GRMHD simulations of a sGRB jet propagating in the dynamical ejecta of a BNS merger. Strong signs of hadronic interactions and related particle cascades were found to be responsible for heavily distorting the Wien spectral energy distribution of photons below the photosphere. This interesting finding called for further investigation to figure out  whether the dissipation scenario, selected based on the jet dynamics, was a peculiar one, conducive to non-thermal distortions. For this purpose, in this work, we explore the shape of the particle spectra for an extended reconnection scenario (where magnetic reconnection is considered to be the main dissipative mechanism, spanning the full radial range under consideration), and a late reconnection scenario where the reconnection region is moved close to the photosphere. We  compare these cases to the benchmark  case presented in~\cite{Rudolph:2023auv}. In addition, we explore  the spectral energy distributions across different jet and cocoon viewing angles. 

The structure of the paper is as follows.  In Sec.~\ref{sec:outflow_model}, we introduce the sGRB jet model as well as the setup of the radiation GRMHD jet simulation and the particle acceleration sites. Section~\ref{sec:MMmodel} outlines the approach adopted to model the radiation and the injection of non-thermal particles. Our results on the non-thermal distortion of the particle spectral energy distributions are presented in Sec.~\ref{sec:results} for photons and neutrinos. Finally, we discuss and summarize our findings in Sec.~\ref{sec:conclusions}. In addition, Appendix~\ref{sec:radiation_modeling} outlines  how to compute  the time evolution of thermal and non-thermal particles, while Appendix~\ref{appendix:jet}  shows snapshots of the particle density distributions and cooling rates for  thermal,  leptonic, and  lepto-hadronic scenarios. 

\section{Outflow model}
\label{sec:outflow_model}
The jet model defines the microphysical conditions in and around the jet as a function of radius. 
After outlining the main features of our benchmark jet simulation, we select  three different viewing angles along which we examine the particle production  later on, 
and identify the  energy dissipation mechanisms and regions. 

\subsection{GRMHD jet simulation}
Our work builds on the GRMHD simulation of a black hole (BH)-powered jet in  homologously expanding  merger ejecta. The simulation has been carried out using the GPU-accelerated code \textsc{h-amr}~\citep{Liska2022}. 
Our GRMHD simulation follows the model $ {\rm \alpha3d5} $ presented in \citet{Gottlieb:2022sis}, albeit with  initial magnetization of $\sigma_0 = 150$. In addition, the jet propagation is followed for  longer times (i.e., up to larger radii). 
The simulation considers a metastable neutron star with lifetime  $t_d = 0.5$~s, which collapses into a BH of mass $ M_{\rm BH} = 3 M_\odot $ and dimensionless spin $ a =0.9375$. 
A torus surrounds  the central compact object (with mass $ M_t = 0.2 M_\odot $ and  characteristic gas-to-magnetic pressure ratio of $1000$ is established during the meta-stable neutron star  phase). The merger ejecta expands homologously with a velocity between $ v_{\rm min} = 0.05 c $ and $ v_{\rm max} = 0.25 c $.
 
The related azimuthally-symmetric baryon density profile is given by 
\begin{equation}\label{eq:density}
	\rho(v_{\rm min}t_d<r<v_{\rm max} t_d,\theta) = \rho_0r^{-3}\left(0.1+{\rm sin}^2\theta\right)\ ,
\end{equation}
where $\theta$ is the polar angle and $\rho_0$ is determined by requiring that the total ejecta mass is $ M_{\rm{ej}} = 0.05\,M_\odot $. 
We refer the interested reader to  \cite{Gottlieb:2022sis} for more details on the simulation setup.
In the regions of interest, adaptive mesh refinement (AMR) is enabled, and the spatial resolution at radius $ r $ is up to $ dr/r = 0.0038 $, depending on the number of AMR levels activated.
The temporal resolution at which the simulation data is sampled is $ 3\,{\rm ms} $.

\begin{figure*}[hbt]
    \centering
    \includegraphics[width = 0.95\textwidth]{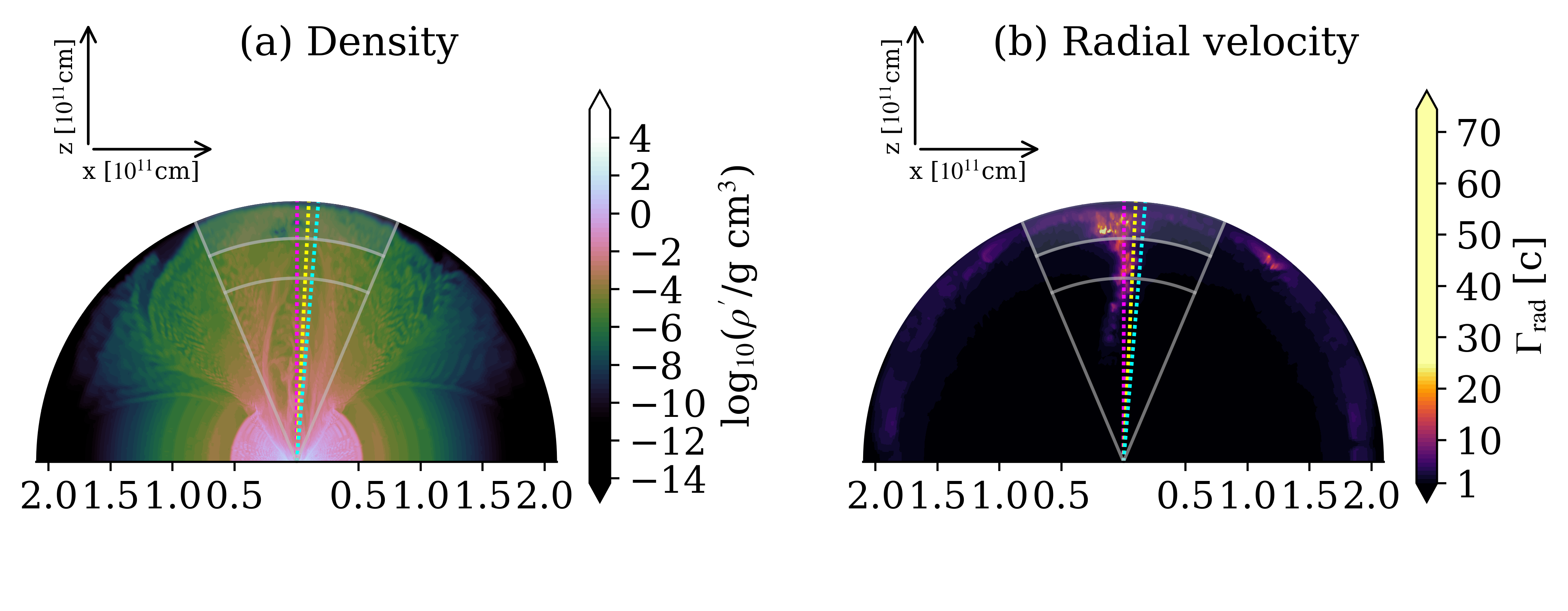}
    \includegraphics[width = 0.95\textwidth]{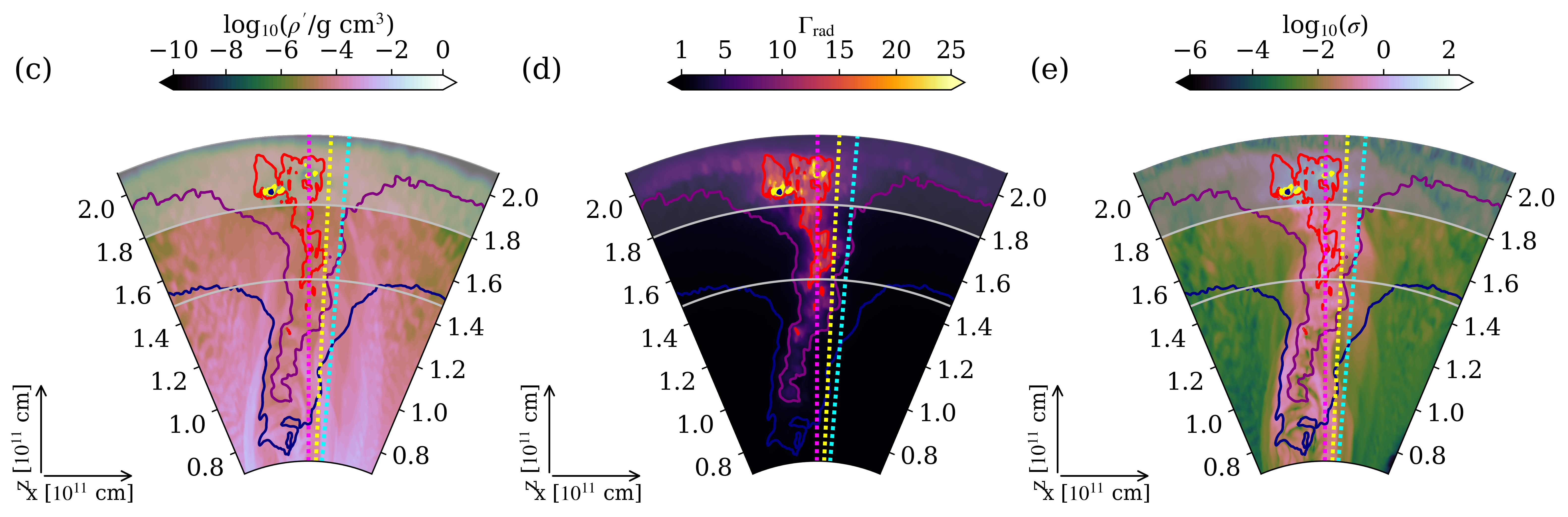}
    \caption{Contour plots of the comoving rest mass  density (left upper and left lower panel, (a) and (c)) and Lorentz factor (right upper and middle lower panel, (b) and (d)) as well as the magnetization (right lower panel, (e)) in  polar coordinates on  the $x$-$z$ plane 
    at $7$~s, 
    when the jet-head is at  $r \approx 2 \times 10^{11}$~cm.  
    The color scales are chosen to enhance the visibility of the relevant structures: Densities below [above] $10^{-8}$ [$10^{4}$]~g/cm$^3$ are marked in black [white], whereas the upper limit for the color scale of the Lorentz factor is set to $25$. The dotted lines  indicate the selected viewing angles $\theta$ equal to $0^\circ$ (``on-axis''), $2.7^\circ$ (``off-axis'') and $4.8^\circ$ (``jet-cocoon interface'').
    The upper panels display the  region between the central engine and the head of the jet, the lower panels represent a zoomed-in version in the angular range marked by light gray radial lines in the upper plot. The lighter shading at the largest radii indicates the simulation region excluded from our analysis.
    We focus our work on the region around the jet between $1.46 \times 10^{11}$~cm and $1.8 \times 10^{11}$~cm, highlighted with  light gray lines.}
    \label{fig:snapshot_7s_x_z}
\end{figure*}

From its launching, we follow the jet evolution for $7$~s, at which time the jet head has reached a radius of $r \simeq 2 \times 10^{11}$~cm. Cuts in the $x$-$z$-plane  of the density and radial profile of the Lorentz factor $\Gamma$ at $7$~s  are plotted in Fig.~\ref{fig:snapshot_7s_x_z}. 
On the left panel, the pink spherical outflow surrounding the central compact object  represents the expanding ejecta. The jet propagation through the ejecta creates a funnel along the $z$-direction, leaving behind a turbulent medium with overall smaller  density. 
The jet region can be identified along the $z$-axis in the right panel of Fig.~\ref{fig:snapshot_7s_x_z}, characterized by a Lorentz factor of $\mathcal{O}(10$--$20)$, surrounded by a slower moving material with $\Gamma \simeq \mathcal{O}(5$--$10)$, and a mildly- to non-relativistic jet-cocoon interface~\citep{Gottlieb2021:structure}. The core radial extension is $\sim 0.6 \times 10^{11}$~cm (extending between $1.5$--$2.1 \times 10^{11}$~cm), corresponding to a BH activity time of $\sim 2$~s. We highlight three viewing angles (cf.~dotted lines): $\theta = 0.1 \, ^\circ$ (along the $z$-axis), $\theta = 2.7\, ^\circ$ and $\theta = 4.8 \, ^\circ$ (from left to right, respectively). These angles  correspond to 
 ``on-axis,'' ``off-axis,'' and ``jet-cocoon interface'' (the latter being labeled as ``cocoon'' in the figures for brevity) emission directions and are identified based on their Lorentz factors.

\begin{figure*}[htb]
    \centering
    \includegraphics[width = \textwidth]{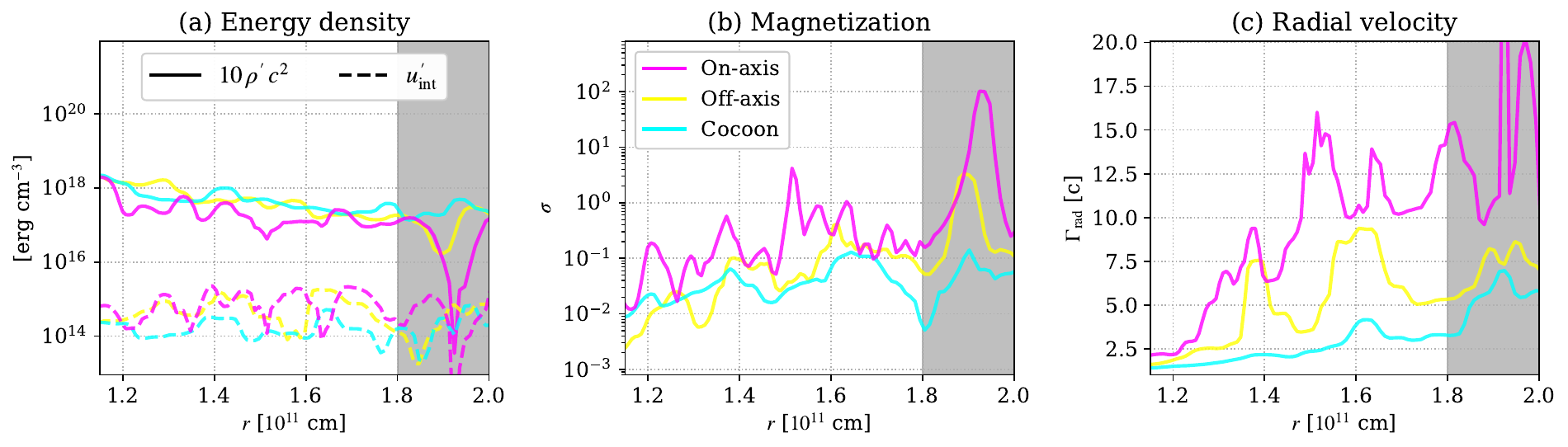}
    \caption{Radial evolution of the characteristic quantities: (a) rest-mass density and internal energy density, (b) magnetization, and (c) radial profile of the  Lorentz factor, from left to right respectively. The quantities are obtained at $7$~s for three characteristic angles $\theta$, marked by  dotted lines in Fig.~\ref{fig:snapshot_7s_x_z}. Due to numerical artifacts, we exclude the region with $r \gtrsim 1.8 \times 10^{11}$~cm (shaded in gray) from our analysis. }
    \label{fig:snapshot_cuts_by_angle}
\end{figure*}

The evolution of the characteristic quantities as  functions of radius (for $\phi = 0$), in the region between $1.2\times 10^{11}$~cm and  $2 \times 10^{11}$~cm, is displayed in Fig.~\ref{fig:snapshot_cuts_by_angle} for the three selected angles. The adopted radial range corresponds to the extension and location of the jet identified in Fig.~\ref{fig:snapshot_7s_x_z}, where the region with $r \gtrsim 1.8 \times 10^{11}$~cm   is shaded in dark gray and will be neglected in the following since  plagued by numerical artifacts.

The left panel of Fig.~\ref{fig:snapshot_cuts_by_angle} shows the rest mass (dashed lines) and internal energy density (solid lines). Going from the on-axis direction to the jet-cocoon interface, the internal energy density tends to decrease whereas the rest mass density increases. While this behavior agrees well with expectations, the values differ less than one order of magnitude between the three viewing angles. Similar to the internal energy density, the magnetization $\sigma$ (middle panel) is the highest on-axis, while being low (and with a smoother profile) within the cocoon; the spread between the three angles reaches up to two orders of magnitude.
Finally, the radial profile of the Lorentz factor 
(shown in the right panel) 
decreases as we move from the on-axis to the jet-cocoon interface viewing directions: while on axis the plasma moves with  $\Gamma_\mrm{rad} \simeq \mathcal{O}(10$--$20)$, the Lorentz factor decreases to $\Gamma_\mrm{rad} \simeq \mathcal{O}(1$--$5)$ in the jet-cocoon interface. Hence, the jet-cocoon interface is characterized by a smooth profile, while stronger fluctuations are visible within the jet.

At the end of the GRMHD simulation, the outflow is still optically thick. Since running the simulation up to the photosphere is numerically challenging,  we  extrapolate the characteristic quantities until larger radii following  the procedure outlined in \citet{Rudolph:2023auv}, 
which includes the calculation of the optical depth to Thomson scatterings. For all radii considered, the optically thin regime is not reached.
In order to do so, first, we define the (radial) region of interest from the last snapshot, which lies  between the  constant-radius lines plotted in light gray in Fig.~\ref{fig:snapshot_7s_x_z}, i.e.~it is between $1.46$ and $1.8 \times 10^{11}$~cm. This region is represented as a single shell (hence, a single radiation zone) moving outward for each of the three selected angles representing the jet/outer jet/jet-cocoon interface. 
Second, we determine the evolution of $\Gamma$, $\sigma$, $\rho$, and $u$. For the first two  quantities, for simplicity, we assume constant values, since their average values change only marginally over the last snapshots of the GRMHD simulation. Only in the jet itself, small regions of  $\sigma \gtrsim 1$ (see Fig.~\ref{fig:snapshot_cuts_by_angle} middle panel) suggest ongoing mild acceleration, which is further supported by the specific enthalpy $h \gtrsim 1$. However, we do not expect a large impact of this on the results. To summarize, we consider the average Lorentz factor within the region of interest extracted at the last snapshot as the final constant Lorentz factor ($\Gamma \equiv \langle \Gamma\rangle_\mrm{7s}$). Similarly, the magnetization equals the average one reached at the last snapshot ($\sigma \equiv \langle \sigma \rangle_\mrm{7s}$).
For the rest-mass density and the internal energy density, we  extrapolate the evolution from the snapshots extracted at  $[5.0, 5.5, 6.0, 6.5, 7.0]$~s as follows: For each fluid cell contained in our region of interest, we fit a power-law to the evolution over snapshots (i.e., $u^\prime \propto r^{-s_u}$ and $\rho^\prime \propto r^{-s_\rho}$). Finally, we take the average values for each angle as the representative ones for the region.

Starting from a radius of $10^{11}$~cm, we follow the evolution of a single radiation zone, which is representative of the region of interest up to $10^{12}$~cm, calculating the rest mass and internal energy density as indicated above. The parameters for the jet, outer jet, and jet-cocoon interface are summarized in Table~\ref{tab:shell_parameters_angles}. Note that this corresponds to the evolution of a  thermal scenario (without taking into account  additional energy dissipation).
\begin{table}[htb]
    \caption{Evolution of the characteristic  parameters (Lorentz boost factor $\Gamma $, magnetization $ \sigma $, spectral indexes, as well as their related standard deviations, for the radial evolution of the internal energy density $ s_u $ and rest-mass density $ s_\rho $) of our shell for the three selected viewing angles. }
    \label{tab:shell_parameters_angles}
    \centering
    \begin{tabular}{c |c c c c c c }
    \toprule
     & $\Gamma$ & $\sigma$ & $s_u$ & $\sigma_{s_u}$ & $s_\rho$ & $\sigma_{s_\rho}$\\ \hline
    On-axis & 12.5 & 0.18 & -3.50 & 1.88& -2.96& 0.86\\
    Off-axis & 6.4 & 0.11 & -2.97 & 0.93 & -3.04& 0.73 \\
    Jet-cocoon interface &  3.1 & 0.06 & -3.09 & 1.04 & -3.34 & 0.42\\
         \bottomrule
    \end{tabular}
\end{table}

\subsection{Particle acceleration sites}
\label{sec:particle_acceleration_sites}

We now introduce the particle acceleration mechanisms and sites identified in our benchmark simulation. As we will show, dissipative processes are crucial for obtaining non-thermal spectral features. We stress that, while the resolution of our jet simulation might not resolve all physical features possibly responsible for particle acceleration in this work we only aim at exploring whether the resultant photon spectra strongly depend on the initial jet conditions.  We do so without coupling the jet  to particle-in-cell simulations, because our kinetic simulations are computationally expensive. Moreover, since we focus on the jet conditions below the photosphere, it is beyond the scope of this work  to reproduce the observed GRB spectra.

\textbf{Magnetic reconnection}.
Magnetic dissipation in GRBs is envisioned through ``striped wind'' scenarios~\citep{Spruit:2000zm, Begue:2016jvr, Beniamini:2017fqh, Giannios:2018nin} relying on regions of inverted polarity produced close to the jet base that enable magnetic reconnection. 
Overall, typically expected length scales are of the order of $\lambda \sim 10^{7}$--$10^{9}$~cm,  below the photosphere and our region of interest. 
For example, the average magnetization of $0.1 \lesssim  \sigma \lesssim 1 $ is reached well below the photosphere, disfavoring magnetic reconnection at $r \gtrsim 10^{11}$~cm \citep[see also][]{Gottlieb:2022sis}. 
However, as discussed in~\citet{Rudolph:2023auv}, our benchmark simulation suggests that a higher magnetization is achieved locally in the jet at $\sim 10^{11}$~cm~\footnote{The average magnetization (weighed with the energy density flux) in the jet decreases throughout the simulation.}, enabling magnetic reconnection at larger radii. 
To investigate this possibility, we mimic the fluctuations of strength $\Delta \sigma$ as a single ``spike'' in $\sigma$ between $r_\mrm{rec}^\mrm{min}$ and  $r_\mrm{rec}^\mrm{max}$. After the build-up of $\sigma$, its subsequent decay with $\mathrm{d}\sigma / \mathrm{d} r \equiv 2 \Delta  \sigma/(r_\mrm{rec}^\mrm{max} - r_\mrm{rec}^\mrm{min})$ leads to a (comoving) rate of magnetic energy dissipation of\footnote{Hereafter, we  denote quantities in the source and plasma comoving frame as $X$ and $X^\prime$, respectively.}:
\begin{equation}
    \frac{\mathrm{d}u_B^\prime}{\mathrm{d}t^\prime} = \rho^\prime c^2 \beta \Gamma c \frac{\mathrm{d}\sigma}{\mathrm{d} r} \, .
\end{equation}
While in \citet{Rudolph:2023auv}, we  followed closely the GRMHD simulation and accounted for magnetic reconnection only within the region covered by the simulation ($1$--$1.6 \times 10^{11}$~cm), we here introduce two additional scenarios in order to explore the potential impact of reconnection at different radii: 
\begin{enumerate}
\item In the \textit{extended reconnection} scenario, the reconnection region spans the full radial range of the jet simulation extending up to $10^{12}$~cm, where the build-up of the magnetization occurs between $ 1$--$5 \times 10^{11}$~cm, followed by a subsequent decay. 
\item In the \textit{late reconnection} scenario the spike keeps the original width but moves out to $9$--$9.6 \times 10^{11}$~cm, which is closer to the photosphere.
\end{enumerate}
As for the off-axis viewing angle and the jet-cocoon interface, we find no indication of a local enhancement of $\sigma$ in the jet simulation and thus conclude that magnetic reconnection  cannot operate outside of the jet core. 

\textbf{Collisionless (sub-)shocks}. Fluctuations in the Lorentz factor of the plasma can introduce collisions between the plasma cells moving at different speeds. For a variability timescale $t_\mrm{var} $ (in the engine frame) the characteristic radius of internal shocks can be estimated as 
\begin{equation}
    r_\mrm{IS} = 2 t_\mrm{var} \Gamma^2 c \, ,
\end{equation}
where $\Gamma$ is the Lorentz factor of the region under investigation. 
If the plasma exhibits variability on $t_\mrm{var}^{\mrm{min}}< t_\mrm{var}< t_\mrm{var}^{\mrm{max}}$, internal shocks should be present at $r_\mrm{IS}^{\mrm{min}}< r < r_\mrm{IS}^{\mrm{max}}$.
From the jet simulation, we identify $50 \, \mrm{ms} \leq t_\mrm{var} \leq 500 \, \mrm{ms}$ for the on-axis viewing angle, 
the off-axis direction and no significant variability is found in the jet-cocoon interface. 

Shocks convert kinetic energy into internal one and the rate of energy dissipation is thus linked to the slow-down of the plasma, $\mathrm{d}\Gamma/\mathrm{d}r$. To estimate the latter, we consider the interaction between the fast (f) and slow (s) parts of the jet located at radii $r_\mathrm{f}$ and $r_\mathrm{s} = r_\mathrm{f} + c t_\mrm{var}$.
After the interaction, the plasma moves at the Lorentz factor
\begin{equation}
    \Gamma_\mathrm{res} = \sqrt{\frac{\rho(r_\mathrm{f}) r_\mathrm{f}^2 \Gamma_\mathrm{f} + \rho(r_\mathrm{s}) r_\mathrm{s}^2 \Gamma_\mathrm{s}}
    {\rho(r_\mathrm{f}) r_\mathrm{f}^2/ \Gamma_\mathrm{f} + \rho(r_\mathrm{s}) r_\mathrm{s}^2/ \Gamma_\mathrm{s} }} \, .
\end{equation}
For a region moving at Lorentz factor $\Gamma$ before the internal shocks,  we obtain the slow down of the plasma due to internal shocks as $\Delta \Gamma =\Gamma - \Gamma_\mathrm{res}$. 
Although in a realistic picture, single shocks may dissipate energy at specific radii, for simplicity we  assume that $\mathrm{d}\Gamma/\mathrm{d}r = \Delta \Gamma / (r_\mrm{IS}^{\mrm{max}} - r_\mrm{IS}^{\mrm{min}}) =$ const. 
This yields a (comoving) rate of energy dissipation by internal shocks
\begin{equation}
    \frac{\mathrm{d}u_\mrm{IS}^\prime}{\mathrm{d}t^\prime} = \rho^\prime c^2 \beta \Gamma c \frac{\mathrm{d}\Gamma}{\mathrm{d} r} \, .
\end{equation}

It has been pointed out repeatedly that acceleration at internal shocks cannot operate efficiently if these occur in the optically thick regime, where the shocks are radiation mediated. 
However, collisionless subshocks capable of particle acceleration may form within RMS,  if the plasma is mildly magnetized  ($\sigma \gtrsim 0.01$) and the following condition is fulfilled
\begin{equation}
    \chi \equiv p_\mrm{th} /p_\mrm{mag} \lesssim 2\ , 
\end{equation}
here $p^\prime_\mrm{mag}= B^{\prime 2} / 8 \pi$ denotes the magnetic pressure and $p^\prime_\mrm{th} = (\hat{\gamma}-1) u^\prime_\mrm{th}$ is the thermal pressure~\citep{Beloborodov:2016jmz}, thus $\chi = 2 u_\mrm{th}^\prime / 3 \sigma \rho^\prime c^2$. We find that this criterion is met both for the on-axis and off-axis viewing angles. Hence, we assume efficient particle acceleration. 
We stress that we only take into account  subshocks in our analysis, and not  RMSs. We neglect the latter since an analytical way to take them into account is not available to date. Future work should aim to  explore the additional effects of RMSs on the photon spectra as well as the interplay between subshocks and the RMSs in which they are embedded in.

\textbf{Magnetic Turbulence}.
Particle acceleration may also occur in MHD turbulence~\citep[see e.g.][]{Lemoine:2021mtv}, possibly in combination with shocks~\citep{Bresci:2023pjx}.
However,  acceleration is not expected to operate efficiently below the photosphere in a collisional plasma, thus is neglected  in the following. 

\begin{figure*}[hbt]
    \centering
    \includegraphics[width = 0.7\textwidth]{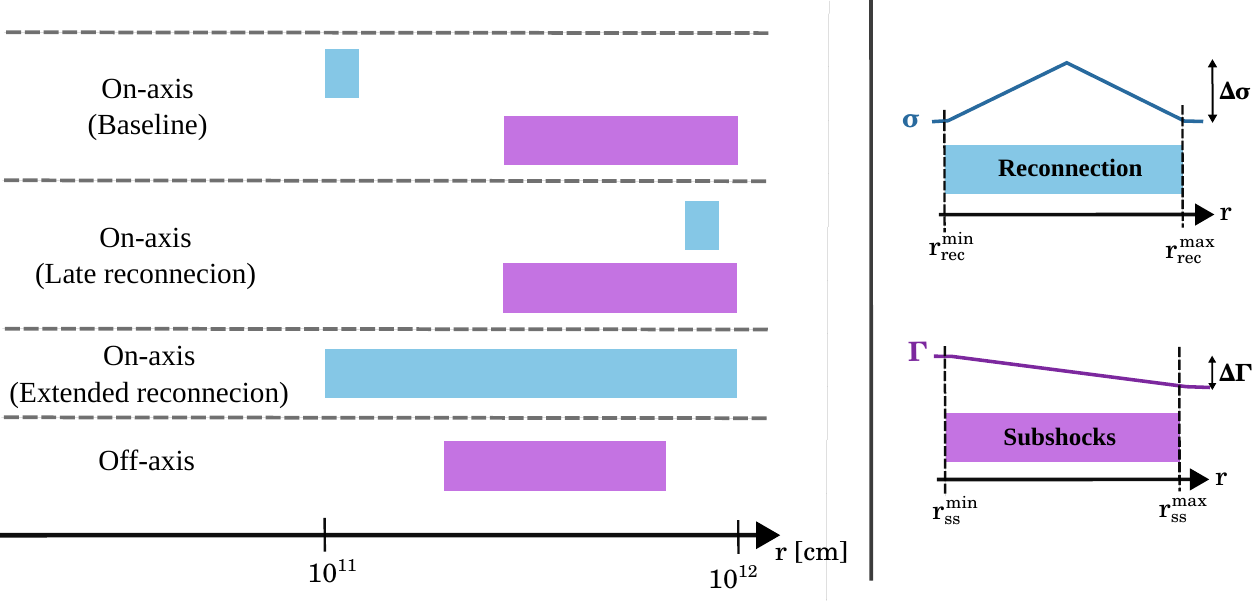}
    \caption{\textit{Left:} Sketch of the radial range of  particle acceleration for the on-axis (benchmark model, late reconnection, and extended reconnection) and off-axis viewing angles, from top to bottom respectively. \textit{Right:} Diagrams illustrating the  radial evolution of $\sigma$ and $\Gamma$ in the regions of magnetic reconnection and collisionless subshocks, respectively.}
    \label{fig:sketch_dissipation_regions}
\end{figure*}

Figure~\ref{fig:sketch_dissipation_regions} summarizes  the investigated dissipation regions for the on- and off-axis viewing angles, as well as the radial profiles of the Lorentz factor $\Gamma$ and magnetization $\sigma$. 
For the on-axis viewing angle, we explore three different scenarios: the \textit{baseline} model investigated in~\citet{Rudolph:2023auv}, the \textit{late reconnection} model where the reconnection region is the same as in the baseline model, but moved outwards to overlap with the subshock region, and the  \textit{extended reconnection} scenario, where the reconnection zone spans the full radial range, but no subshocks are included. This is complemented by the off-axis scenario, in which only sub-shocks operate.
We remind the reader that particle acceleration is inefficient in the jet-cocoon interface.

Figure~\ref{fig:sketch_dissipation_regions} further illustrates that, for magnetic reconnection, we simplify the evolution and assume a single spike in $\sigma$, where particle acceleration  takes place in the second half. For subshocks, a steady decrease of the Lorentz factor of the outflow is assumed to be due to the transformation of kinetic energy into internal.
The  assumptions on the radial range as well as  $\Delta \sigma$ and $\Delta \Gamma$ are summarized in Table~\ref{tab:dissipative_scenarios}.
\begin{table*}[]
    \caption{Summary of the particle acceleration regions and variations in $\Gamma$ as well as $\sigma$ for the collisionless subshocks and magnetic reconnection scenarios for the three selected viewing angles, including late and extended reconnection for the on-axis case. }
    \label{tab:dissipative_scenarios}
    \centering
    \begin{tabular}{c |c c c | c c c}
    \toprule
              &  \multicolumn{3}{c |}{Subshocks} & \multicolumn{3}{c}{Magnetic reconnection} \\ \hline
         Model  & $r^\mrm{min}_\mrm{ss}$ [cm] & $r^\mrm{max}_\mrm{ss}$ [cm] & $\Delta \Gamma$ & $r^\mrm{min}_\mrm{rec}$ [cm] & $r^\mrm{max}_\mrm{rec}$ [cm] & $\Delta \sigma$ \\ \hline
         On-axis (Baseline) & $4.7 \times 10^{11}$ & $1.0 \times 10^{12}$ & 0.5 & $1 \times 10^{11}$ & $1.6 \times 10^{11}$ & 7.5 \\
         On-axis (Late reconnection) & $4.7 \times 10^{11}$ & $1.0 \times 10^{12}$ & 0.5 & $9 \times 10^{11}$ & $9.6 \times 10^{11}$ & 7.5 \\
         On-axis (Extended reconnection) & -- & -- & -- & $1 \times 10^{11}$ & $1 \times 10^{12}$ & 7.5 \\ \hline
         Off-axis & $4.22 \times 10^{11}$ & $8.45 \times 10^{11}$ & 0.47 & -- & -- & -- \\ 
         Jet-cocoon interface  & -- & -- & -- & -- & -- & -- \\
         \bottomrule
    \end{tabular}
\end{table*}

\begin{figure*}[htb]
    \centering
    \includegraphics[width = 0.93\textwidth]{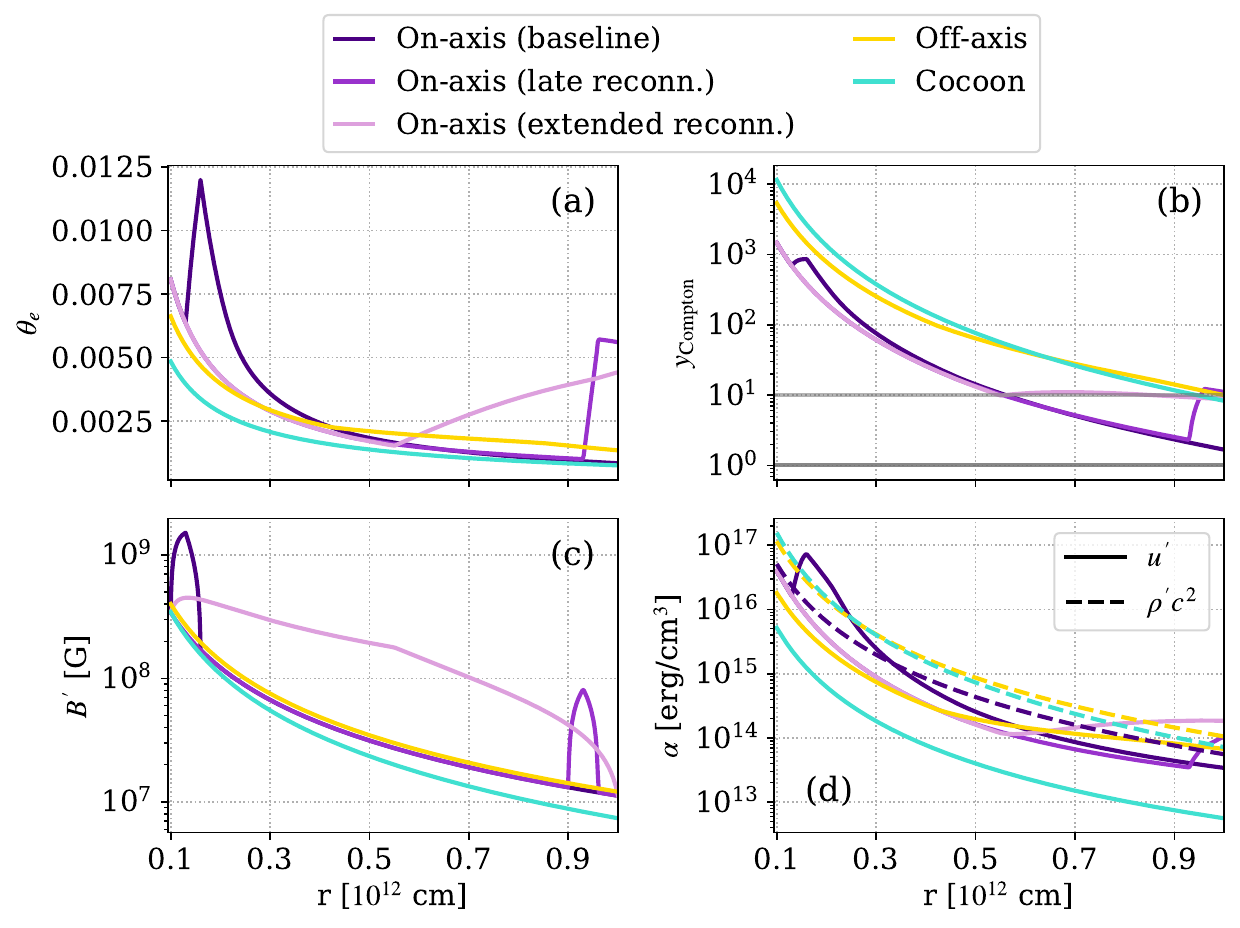}
    \caption{Radial evolution of characteristic quantities for the representative radiation zone: purple lines refer to  on-axis scenarios (baseline, late reconnection, and extended reconnection are plotted with different shades), the off-axis (jet-cocoon interface) scenario is plotted in yellow (cyan). We plot (a) the dimensionless electron thermal temperature ($\theta_e = T_e k_B /m_e c^2$), (b) the Compton parameter ($y_\mrm{compton} = 4 \theta_e \sigma_T n_e r /2 \beta \Gamma c $), (c) the comoving magnetic field $B^\prime$, and (d) the comoving energy densities associated with the internal energy ($u^\prime$) and rest mass ($\rho^\prime c^2$) from top left to bottom right, respectively. }
    \label{fig:evolution_5_models}
\end{figure*}

Figure~\ref{fig:evolution_5_models} illustrates the radial evolution of the characteristic quantities relevant for radiation modeling (thermal electron temperature, magnetic field strength, energy, and rest mass densities, the Compton  parameter), obtained by combining the extrapolated thermal behavior with the identified energy dissipation regions for the five cases summarized in Table~\ref{tab:dissipative_scenarios}. 

The left upper panel shows the thermal electron temperature, $\theta_e = k_B T_e/(m_e c^2)$. While for thermal evolution, this temperature has a simple power-law decay (e.g., in the jet-cocoon interface, plotted in cyan), energy dissipation heats the thermal populations, with  resulting spikes in the temperature. This is most pronounced for the three on-axis scenarios where magnetic dissipation is efficient. On the other hand, internal sub-shocks do not heat the population  efficiently. The upper right plot shows the Compton  parameter defined as 
    $y_\mathrm{Compton} = 4 \theta_e \sigma_T n_e t^\prime_\mrm{dyn}$.
For all models, $y_\mathrm{Compton} > 1$ within our region of interest, implying that the outflow is optically thick.
The magnetic field profile (bottom left panel, between $10^7$--$10^9$~G with little variance across the three viewing angles) carries the signs of the assumed simplified spikes for the three on-axis scenarios.
The rest-mass density (bottom right panel) is unaffected by the dissipative scenarios and follows a simple power-law decay, the energy density $u^\prime$ (bottom right panel) is enhanced because of plasma heating in dissipative regions (both through magnetic reconnection and internal subshocks).

\section{Multi-messenger emission modelling}
\label{sec:MMmodel}
In the following, we introduce the  method employed to model the spectral energy distributions  of  different particle species, including the dissipation mechanisms. This approach  allows to  compute the multi-messenger signals expected along the three selected viewing angles and dissipation scenarios.

\subsection{Radiation modelling}

In a thermal plasma, electrons and photons interact through Compton scattering, double Compton, thermal bremsstrahlung, as well as cyclotron emission and absorption. If particle acceleration is efficient, leptonic and hadronic emission processes contribute to shaping the spectral energy distributions. Charged particles emit synchrotron radiation in the presence of a magnetic field and undergo inverse Compton scattering processes. Photon-photon annihilation and photo-hadronic pair production lead to  electron-positron pairs, and the resulting secondary pairs can affect the photon distribution through synchrotron and inverse Compton processes. Moreover, proton-proton interactions and photo-pion production yield pions that then decay into photons, muons, electrons, and neutrinos. 
In order to take into account all these processes and shape the spectral energy distributions, at each radius, we solve the coupled partial differential equations (PDEs) of electrons, positrons, protons, neutrons, pions, muons, neutrinos, and photons. 

The PDEs take the form of  transport equations:
\begin{align}
    \delta_t n(E) = & \frac{1}{A(E)} \delta_E \left( D(E)\delta_E n(E) + a(E) n(E) \right) \\
                    & + \epsilon(E) - \alpha(E) n(E) \, ,
\end{align}
where $\delta / \delta_x = \delta_x$, $n$ is the particle number (or occupation number, for momentum space), $E$ is the particle energy, and $A(E) = E^2$, if particles are treated in momentum space and $A(E) =1$ otherwise. The remaining factors account for diffusion [$D(E)$], advection/cooling [$a(E)$], source/injection [$\epsilon(E)$], and sink/escape [$\alpha(E)$] terms. 

The evolution of non-thermal electrons, positrons, protons, neutrons, pions, muons and neutrinos is tracked relying on a modified version of the AM$^3$ code~\citep{Gao:2016uld}, which includes all relevant non-thermal processes such as synchrotron, inverse Compton, photo-pion production, photo-pair production, proton-proton interaction, $\gamma \gamma $-annihilation and adiabatic cooling. Note that inverse Compton scatterings are  computed on the full photon distribution and hence  include synchrotron-self-Compton contributions.
Secondary particles are included self-consistently, undergoing the same interactions as primary particles. 
Photons are treated in momentum-space with the Kompaneets equation~\citep{1957:Kompaneets} extended to include cooling, absorption and emission terms, see Appendix~\ref{sec:radiation_modeling} for more details.
The populations of thermal electrons, neutrons, and protons are described solely by their density and temperature. We have verified that, for all radii and scenarios considered, a Maxwellian distribution can  be safely assumed, since the relaxation time due to Coulomb collisions is shorter than the dynamical timescale.

\subsection{Injection of non-thermal populations}

The non-thermal particle populations of the species $i$ are injected in the form of a power-law with exponential cutoffs below the minimum Lorentz factor $\gamma^\prime_\mrm{i, min}$ and above the maximum Lorentz factor $\gamma^\prime_\mrm{i, max}$ with power-law slope $p_\mrm{i}$:
\begin{equation}
\frac{\mrm{d}n^\prime_\mrm{i}}{\mrm{d}\gamma_\mrm{i}\mrm{d}t^\prime}= A_i \gamma_\mrm{i}^{\prime -p_\mrm{i}} e^{-\gamma^\prime_\mrm{i, min}/ \gamma_\mrm{i}^\prime}e^{-\gamma_\mrm{i}^\prime/ \gamma^\prime_\mrm{i, max}} \, .
\end{equation}
The normalization $A_\mrm{i}$ is calculated from the dissipation rate $\mrm{d} u^\prime/\mrm{d}t^\prime $:
\begin{equation}
   \epsilon_\mrm{i} \frac{\mrm{d} u^\prime }{\mrm{d}t^\prime } \mbeq \int \mrm{d}\gamma^\prime_\mrm{i} \: \gamma^\prime_\mrm{i} m_\mrm{i} c^2 \frac{\mrm{d}n^\prime_\mrm{i}}{\mrm{d}\gamma^\prime_\mrm{i}\mrm{d}t^\prime} \, .
\end{equation}
Here we have introduced the microphysics parameter $\epsilon_\mrm{i}$ which specifies the fraction of energy transferred to the species $i$. In addition, we assume the following for magnetic reconnection and collisionless subshocks: 
 
\textbf{Magnetic reconnection.}
The maximum Lorentz factor of electrons is given by $\gamma_\mrm{e, max}^\prime \sim 5 \sigma m_p / m_e$ \citep[as shown without and with cooling in][]{Werner:2016fxe, Kagan:2017rls}. Following~\citet{Kagan:2017rls}, the minimum Lorentz factor is $\gamma_\mrm{e, min}^\prime = \gamma_\mrm{e, max}^\prime/40$.

For ions, we differentiate between the non-relativistic and relativistic regimes. For $\sigma \leq 10$, the maximum kinetic energy never exceeds $E_\mrm{p, max}^\prime = 3$~MeV~\citep{Werner:2016fxe}. The contribution to non-thermal particle spectra by protons with kinetic energy below $3$~MeV is expected to be negligible, so we neglect the acceleration of protons for $\sigma \leq 10$. On the other hand, in the ultra-relativistic regime, ion spectra resemble the ones of electrons when expressed in particle kinetic energy. This implies that the minimum Lorentz factor is given by $\gamma_\mrm{p, min}^\prime = [(\gamma_\mrm{e, min}^\prime -1) m_\mrm{e}/ m_\mrm{p}] +1 $, whereas the maximum Lorentz factor is $\gamma_\mrm{p, max}^\prime = [(\gamma_\mrm{e, max}^\prime -1) m_\mrm{e}/ m_\mrm{p}] +1$. 
Note that we assume that particles may be accelerated above the synchrotron burnoff limit due to acceleration along the electric field lines in regions of low magnetic fields~\citep[among others][]{Cerutti:2013mma, Kagan:2016rze, Kagan:2017rls}. 

The slope $p_\mrm{e} = p_\mrm{p} = p$ is parametrized following \citet{Werner:2016fxe}:
\begin{equation}
    p = 1.9 + 0.7 / \sqrt{\sigma} \, .
\end{equation}
Note that this $p$ is also approximately compatible with \citet{Zhang:2023lvw}, yielding a power-law slope of $p \sim 1.9 $ for large $\sigma$.

We assume that half of the dissipated energy is converted into internal energy ($\epsilon_\mrm{th} = 1/2$) and the other half is split between accelerated protons and electrons \citep{Sironi:2015eoa}. Electrons receive a fraction $q_\mrm{e}$ of the latter \citep{Werner:2016fxe}: 
\begin{equation}
    q_\mrm{e} = \frac{1}{4} \left( 1 + \sqrt{\frac{\sigma / 5 }{2 + \sigma/5}} \right)\ .
\end{equation}
This choice approaches $q_\mrm{e}= 1/2$ for large $\sigma$, consistent with the equal distribution of energy between electrons and ions at high $\sigma$ found e.g.~in~\cite{Sironi:2014jfa}. 
Thus, electrons receive a fraction of energy $\epsilon_\mrm{e} =  q_\mrm{e}/2$ and protons  $\epsilon_\mrm{p} = (1-  q_\mrm{e})/2$. 
For $\sigma < 10$, the heating of protons instead causes heating of the thermal populations and $\epsilon_\mrm{p}$ is added to $\epsilon_\mrm{th}$. 

The literature quoted above  focuses on  collisionless plasma. However, in the dense environment below the photosphere radiation pressure, radiative cooling, and other effects such as pair creation may play an important role. This regime is unfortunately poorly explored~\citep[see e.g.][for a discussion on the relevant processes]{Uzdensky:2011ka}. For example, the magnetic reconnection rate may be significantly lower in a pair-loaded plasma~\citep{Hakobyan:2018fwg}. Future work including magnetic reconnection in a collisional plasma may shed  light on the accelerated particle distributions in this regime.

\begin{figure*}[hbt]
    \centering
    \includegraphics[width = 0.95\textwidth]{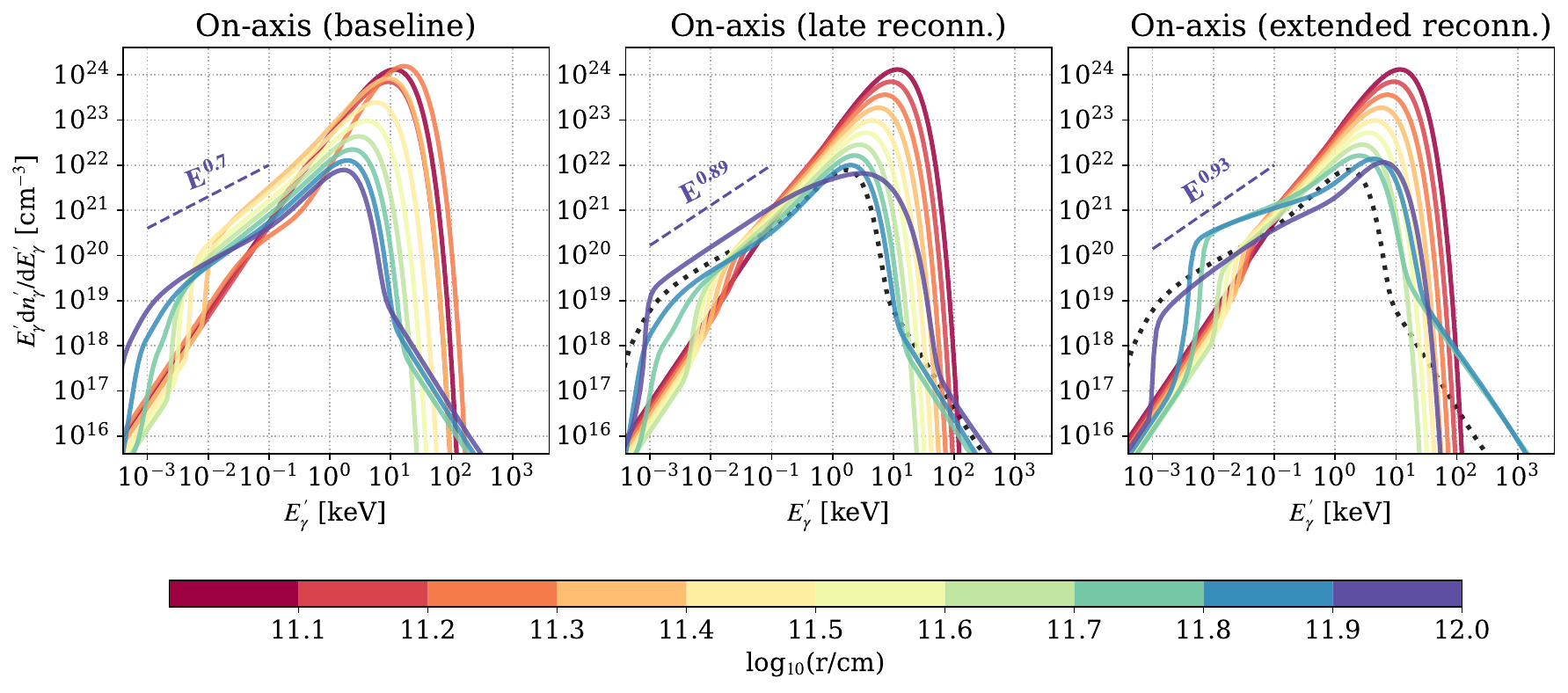} 
    \caption{Evolution of the comoving photon spectral energy distribution for our three on-axis dissipation scenarios (baseline, late reconnection, and extended reconnection, from left to right, respectively). Magnetic reconnection is responsible  for the acceleration of electrons, while protons and electrons are accelerated at the collisionless sbushocks and in the 
    superposition between the reconnection and subshock regions. As a result of plasma expansion, the photon density tends to decrease as the radius increases. In the middle and right plots, the dotted line represents the final photon spectrum of the Baseline model. A power-law fit is provided for the spectral distribution extracted at the final snapshot to facilitate the comparison among the different scenarios. For all cases, the photon spectrum is highly distorted with respect to the initial thermal one, with a variation of $\sim20\%$ in the spectral index fitting the final photon spectrum across the different scenarios. }
    \label{fig:compare_dissipative_cases}
\end{figure*}

\textbf{Collisionless subshocks.} In  collisionless subshocks, acceleration can be assumed to be similar to the case of  mildly relativistic, mildly magnetized shocks in the optically thin regime explored in~\citet{Crumley:2018kvf}: Protons and electrons receive a fraction of $\epsilon_\mrm{p} = 0.1$ and $\epsilon_\mrm{e} = 5 \times 10^{-4}$ of the dissipated energy, respectively. The remaining energy goes into thermal heating. 
Protons are accelerated to a power-law with index $p_\mrm{p} = 2$ above $E_\mathrm{min, p}^\prime = 1$~GeV, whereas the accelerated electrons follow a power-law with index $p_\mrm{e} = 2.2 $ above $E_\mathrm{min, e}^\prime \simeq 2 m_\mrm{p} c^2 $. 

The maximum energy is obtained by balancing the energy losses with the acceleration rate. For a particle of mass $m$ and Lorentz factor $\gamma$, the latter reads
\begin{equation}
    t^{\prime -1}_\mathrm{acc} (\gamma^\prime) = \frac{\eta e B^\prime}{m c \gamma^\prime} \ , 
\end{equation}
where we assume   $\eta = 0.1$~\citep{Globus:2014fka} as acceleration efficiency.

\section{Particle spectral energy distributions}
\label{sec:results}
In this section, we investigate how the photon spectral energy distribution changes according to the  dissipation scenarios and the viewing angle. We also look into the variations of the neutrino signal. 

\subsection{Photon distributions resulting from different dissipation scenarios}

Figure~\ref{fig:compare_dissipative_cases} shows the evolution of the comoving spectral energy distribution of photons for the on-axis viewing angle, for our benchmark dissipation scenario investigated in~\cite{Rudolph:2023auv}, as well as the late reconnection and the extended reconnection scenarios from left to right, respectively. All cases exhibit a robust non-thermal modification of the spectral energy distributions. This result holds for the extended reconnection scenario, where only electrons are accelerated, as well as for the late reconnection and our baseline models where both electrons and protons are simultaneously accelerated at subshocks, while magnetic fields  accelerate electrons either outside or within the subshock region. 
Scrutinizing the processes responsible for photon emission (see Appendix~\ref{appendix:jet}), we find that in both leptonic and lepto-hadronic scenarios, the main additional photon source in non-thermal scenarios is electron synchrotron emission. Here, the electrons can either be primaries (in leptonic cases) or arise as secondaries from hadronic interactions. In the second case, proton-proton interactions are the dominant production channel. The low proton energies almost completely inhibit photo-pion production, while photo-pair production is efficient at the highest proton energies. 
Below the spectral peak, the spectrum is shaped by the Comptonization of these seed photons. In the Wien zone, photons cannot return to thermal equilibrium even after the injection of additional seed photons has ceased, and  the spectral shape is expected to be maintained when the energy dissipation and subsequent particle acceleration have stopped. In contrast, we find that the high-energy tail breaks down fast.
The photon peak energy is dictated by the thermal electron population, and hence increases as the latter is heated in regions of energy dissipation.
The photon peak energy varies mildly among the different scenarios that we explore, being $\mathcal{O}(1$--$10)$~keV in the comoving frame. Considering that the Lorentz factor of the outflow is $\Gamma \simeq 12.5$, this implies  observed peak energies of $\mathcal{O}(10$--$100)$~keV. Such energies are below the peak energies observed for  GRBs, however may be reconciliated with observations with a higher  Lorentz factor and taking into account the  jet evolution and related particle production beyond the photosphere.

While the spectral index characterizing the photon distribution can change up to $\sim20\%$ across the different models, it is interesting that the resulting photon spectrum mildly depends on the specific dissipation scenario; hence, the photon spectrum is affected by  the jet properties, rather than the specific process responsible for energy dissipation. This finding suggests that, most likely, observations may be compatible with different dissipation scenarios (and both leptonic as well as lepto-hadronic scenarios), but in all cases, the region of energy dissipation must be located close to the photosphere, where more target particles are available.  

\subsection{Photon and neutrino distributions resulting from different emission angles}

\begin{figure*}[hbt]
    \centering
    \includegraphics[width = 0.95\textwidth]{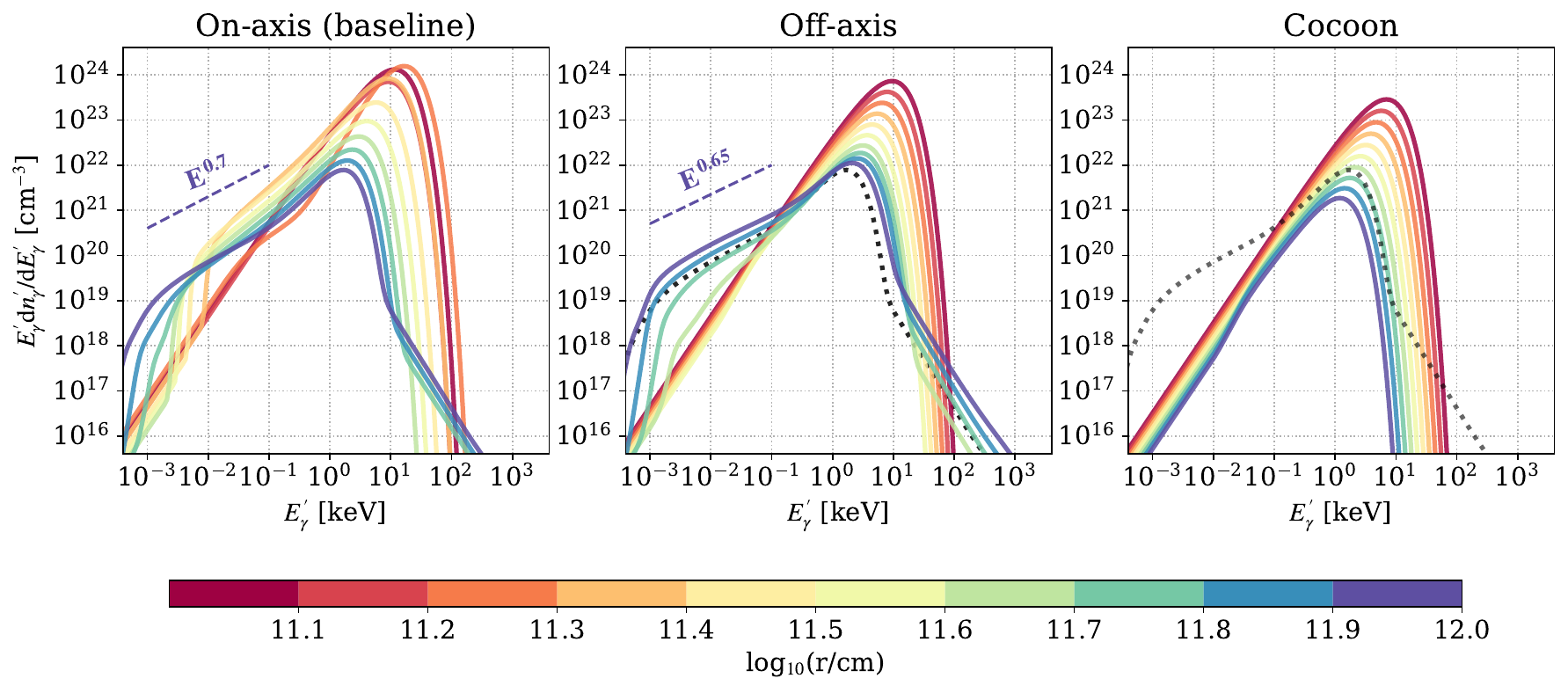} 
    \caption{Same as Fig.~\ref{fig:compare_dissipative_cases}, but for our three selected viewing angles: on-axis, off-axis and jet-cocoon interface. While the spectral energy distribution from the jet-cocoon interface is quasi-thermal, the spectral distributions from different emission angles across the jet (on- and off-axis) are highly non-thermal and relatively similar to each other, showing a variation of the spectral index of $\sim5\%$. 
    }
    \label{fig:compare_dissipative_cases1}
\end{figure*}
Figure~\ref{fig:compare_dissipative_cases1} displays the final spectral energy distributions of photons for the on-axis viewing angle [i.e., our benchmark scenario investigated in~\cite{Rudolph:2023auv}; left], the off-axis viewing angle (with a dissipation model identical to the baseline one; middle), and across the jet-cocoon interface (right). Particle acceleration is inefficient in the jet-cocoon interface due to weaker magnetization and smooth radial profile (cf.~Fig.~\ref{fig:snapshot_cuts_by_angle}), both of which prevent the formation of collisionless subshocks. As a consequence, the final spectrum is quasi-termal, with the only modification due to a decrease in the overall number density of particles because of plasma expansion. On the other hand, the on-axis and off-axis spectral distributions are distorted by non-thermal processes.
Again, in both scenarios, the main additional photon source is electron synchrotron emission, which below the peak is subject to efficient Comptonization. Interestingly, although for the off-axis scenario, energy dissipation has ceased well below the final radius of our simulation, high-energy electrons and their synchrotron emission are still present at the end of the simulation, forming a high-energy tail in the photon spectrum (see Appendix~\ref{appendix:jet} for details). In this sense, (lepto-)hadronic scenarios can sustain non-thermal high-energy signatures above the region of energy dissipation.
As the jet characteristic properties are similar across the jet, the final photon distributions from the on-axis and off-axis cases are comparable, with a difference in the  spectral index of $\sim5\%$.

The observed photon spectrum would be given by the projection along the line of sight of the spectra emitted across different directions pointing towards the observer. 
However, if  observed GRBs should have spectral energy distributions comparable to the ones we observe in this simulation,  one should not expect a large variation in the spectral properties between on- or off-axis observers. The thermal spectrum in the jet-cocoon interface is expected to be a generic feature.

\begin{figure}[hbt]
    \centering
    \includegraphics[width = 0.95\linewidth]{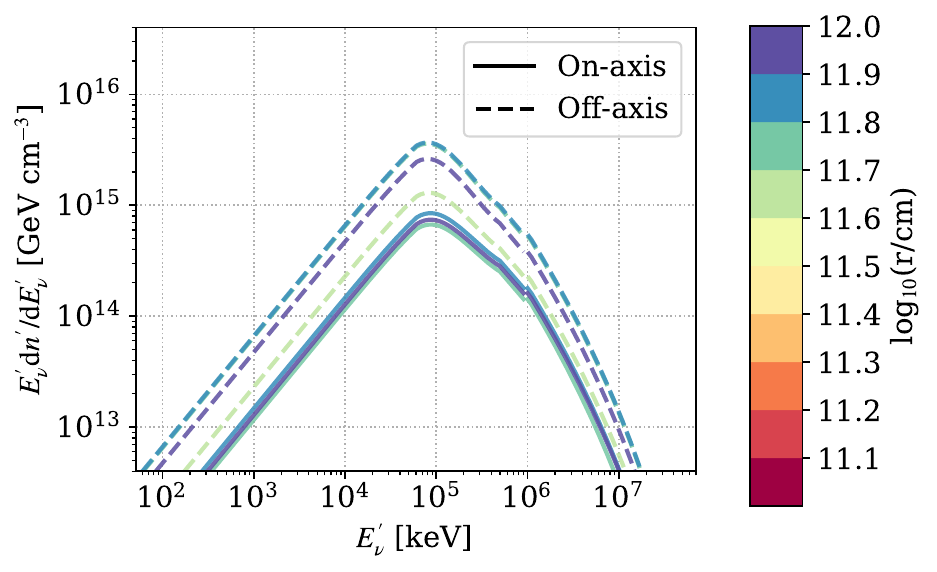} 
    \caption{Evolution of the comoving spectral energy distribution of neutrinos (summed over all flavors) for the on-axis (baseline model) and off-axis scenarios. 
    The late reconnection scenario is not shown, since the neutrino spectral energy distribution is very similar to the baseline one, while  no neutrinos are produced in the extended reconnection scenario. The same color scale  adopted for Fig.~\ref{fig:compare_dissipative_cases} is used here. }
    \label{fig:neutrinos}
\end{figure}
The evolution of the associated comoving neutrino spectral energy distribution from the on- and off-axis scenarios is displayed in Fig.~\ref{fig:neutrinos}. 
In this case, we have focused on our baseline model, we note that the proton density is larger off-axis, leading to a more efficient neutrino production through hadronic interactions. 
For the  late reconnection scenario, a neutrino signal comparable to the baseline one is found (results not shown here); the spectral shape of the neutrino spectrum is similar for the on-axis and off-axis scenarios. On the other hand, for the extended reconnection scenario, no neutrinos are produced.

\section{Conclusions}
\label{sec:conclusions}
Building on a state-of-the-art GRMHD simulation of a short GRB, we have followed the propagation of a relativistic jet with initial magnetization $\sigma_0=150$ for $7$~s in massive ejecta and investigated the photon and neutrino signals emerging from the photosphere. We have computed the  evolution of thermal and non-thermal particles in a  plasma shell representative of the jet between $10^{11}$~cm and $10^{12}$~cm. 

Our baseline scenario is based on the one considered in~\cite{Rudolph:2023auv}, where a sudden enhancement in $\sigma$ enables magnetic reconnection. In this case, neutrinos cannot be produced, because  protons cannot be accelerated to non-thermal energies efficiently. On the other hand, the radiation of non-thermal leptons introduces  non-thermal spectral features. Later, the variability of the Lorentz factor leads to collisionless subshocks which gives rise to non-thermal radiation that significantly alters the thermal spectrum inducing cascades of secondary particles. Since for collisionless sub-shocks, protons are co-accelerated with electrons, neutrino signatures are produced in this second dissipation region.

We have also considered the case  of late reconnection with subshocks and magnetic reconnection (the latter occurring towards the end of the acceleration region), and another scenario with on-axis extended reconnection, where reconnection occurs throughout the acceleration region, but no collisionless sub-shocks are present (and hence no proton acceleration occurs). In these cases, we also find robust non-thermal modifications of the spectral energy distributions, with variations  up to $\sim20\%$ in the spectral index characterizing the energy distribution--independent of whether  leptonic or lepto-hadronic processes were dominating.

Similar conclusions hold for the photon spectral energy distribution for on- and off-axis viewing angles (with a variation in the spectral index of $\sim5\%$). In the cocoon (outside of the jet opening angle),  particle acceleration is inefficient due to unfavorable conditions for collisionless subshocks and magnetic reconnection. The corresponding neutrino spectral energy distribution is similar for the baseline and late reconnection scenarios and is larger for off-axis emission compared to on-axis. We find that neutrinos are not produced in the extended reconnection model. 

The mild modifications of the spectral energy distributions that we find due to the different dissipation scenarios stem from the calculation of the same GRMHD jet simulation. Future work should aim to investigate the variations in the shape of the photon spectral energy distributions for a range of jet models with different characteristic properties and emission. Especially scenarios with a higher Lorentz factor within the jet may be more representative of observed  GRBs. On the other hand, the mild dependence on the dissipation scenario suggests that GRB observations may be compatible with a range of dissipation scenarios as long as the dissipation region is close to the photosphere. 
In addition, future work may account for additional effects, such as RMSs. While not relevant for particle acceleration, energy dissipation at RMSs may still alter the number of photons available, which in a regime of efficient Comptonization can have a considerable impact on the spectral shape. Of particular interest may be the relationship between collisionless subshocks and the larger RMSs they are embedded in, which remains poorly explored.

We  stress that the non-thermal distortions computed in this paper refer to the subphotospheric region. If energy dissipation should take place close to/above the photosphere or in an extended region,  a larger impact of the non-thermal dissipation mechanisms on the photon distribution should be expected. Further, energy dissipation in the optically thin regime will be easier to trace observationally through timing information.
Future work may further address the anisotropy and  
inhomogeneity of the medium, which have not been accounted for in this work. They may be informed through additional considerations of particle acceleration in optically thick media, which as of today remains poorly understood.

\section*{Acknowledgments}
This project has received funding from the Villum Foundation (Project No.~37358), the Carlsberg Foundation (CF18-0183), and the Deutsche Forschungsgemeinschaft through Sonderforschungsbereich SFB 1258 ``Neutrinos and Dark Matter in Astro- and Particle Physics'' (NDM). OG is supported by Flatiron Research Fellowship and acknowledges support by Fermi Cycle 14 Guest Investigator program 80NSSC22K0031 as well as NSF grant AST-2107839. Part of this work was performed at Aspen Center for Physics, which is supported by National Science Foundation grant PHY-2210452.
The Tycho supercomputer hosted at the SCIENCE HPC Center at the University of Copenhagen was used for supporting the numerical simulations presented in this work. An award of computer time was provided by the ASCR Leadership Computing Challenge (ALCC), Innovative and Novel Computational Impact on Theory and Experiment (INCITE), and OLCF Director's Discretionary Allocation programs under award PHY129. This research used resources of the National Energy Research Scientific Computing Center, a DOE Office of Science User Facility supported by the Office of Science of the U.S. Department of Energy under Contract No.~DE-AC02-05CH11231 using NERSC award ALCC-ERCAP0022634.



\appendix

\section{Radiation modeling methods}
\label{sec:radiation_modeling}
In this appendix we describe how we calculate the time evolution of thermal and non-thermal particles. Note that for convenience we change the notation for the reference frames: The plasma comoving frame and the source frame will be denoted as $X$ and $\tilde{X}$, respectively. 
\subsection{Photon spectra: The (extended) Kompaneets Equation}

The evolution of a photon spectrum scattering on a Maxwellian electron distribution can (without explicitly solving the equation for the electron distribution) be described by the Kompaneets equation~\citep{1957:Kompaneets}:
\begin{equation}
    \frac{\partial f}{\partial t} = \frac{\sigma_\mrm{T} n_\mrm{e} c}{x^2}\frac{\partial}{\partial x} x^4 \left( \theta_e \frac{\partial f}{\partial x} + f^2 + f \right) \, .
    \label{eq:kompaneets}
\end{equation}
Here, $f$ is the occupation number of photons (or phase space density) per unit volume, $n_\mrm{e}$ the electron number density, $\theta_e = k_\mrm{B} T_\mrm{e} / m_\mrm{e} c^2$ the dimensionless electron temperature and $x = h \nu / m_\mrm{e} c^2$ the dimensionless photon energy. From the occupation number the photon spectrum is obtained as 
\begin{equation}
    n_\gamma(x) = \frac{8 \pi (m_\mrm{e} c^2)^2 }{(ch)^3} x^2 f(x) \, .
\end{equation}
The total number and energy of photons are calculated as
\begin{align}
n_\gamma &=  \frac{8 \pi (m_\mrm{e} c^2)^3 }{(ch)^3} \int_0^\infty x^2 f \mrm{d} x \\ 
u_\gamma & =  \frac{8 \pi (m_\mrm{e} c^2)^4 }{(ch)^3} \int_0^\infty x^3 f \mrm{d} x \, .
\label{eq:photon_energydensity}
\end{align}
The characteristic time between two scatterings is given by $t_\mrm{c} = 1/ \sigma_\mrm{T} n_\mrm{e} c$. 

We extend the Kompaneets equation by source, sink and cooling terms as follows: 
\begin{align}
    \frac{\partial f(x)}{\partial t}  = \frac{1}{x^2}\frac{\partial}{\partial x}\left[   x^4 \sigma_\mrm{T} n_\mrm{e} c \left(\theta_e \frac{\partial f(x)}{\partial x} + f(x)^2 + f(x) \right)  -  x^2 a_\mrm{exp} (x)  f(x) \right]
    + \epsilon(x) - \tau^{-1} (x) f (x)
    \label{eq:kompaneets_extended}
\end{align}
In addition to cooling and density dilution due to the plasma expansion,   absorption (expressed through the absorption timescale $\tau$) and source terms (expressed as $\epsilon (x) $) in the Kompaneets equation are determined by three radiative processes: bremsstrahlung, thermal cyclotron, and double Compton. The overall cooling and absorption/emission terms are considered to be the sum of the terms due to single processes, namely:

\begin{align}
\tau^{-1}(x) & = \tau_\mrm{FF}^{-1} (x) + \tau_\mrm{CY}^{-1} \, ,
 + \tau_\mrm{DC}^{-1} + \tau^{-1}_\mrm{exp} \\
\epsilon(x) & = \epsilon_\mrm{FF} (x) + \epsilon_\mrm{CY}
 + \epsilon_\mrm{DC} \, ,\\
 a(x) &= a_\mrm{exp} .
\end{align}

Here ``FF'' stands for Bremsstrahlung, ``CY'' for cyclotron, ``DC'' for double Compton, and ``exp'' represents the expansion terms. The exact derivations of the terms will be introduced below.
Below we give the absorption coefficients for these processes, following \citet{Vurm:2012be} for their implementation  \citep[which is consistent with][]{Begue:2014kxa, 1986:RybickiLightman}. 
We assume that the emissivity ($j_x$) and the absorption coefficient ($\alpha_x$) can be related to each other by Kirchhoff law, valid for an electron distribution in thermal equilibrium:
\begin{equation}
 \frac{j_x}{\alpha_x} \equiv B_x = \frac{2 x^3 (m_\mrm{e} c^2)^3 / (hc)^2}{\mrm{exp}(x/\theta_\mrm{e}) -1 } \, .
 \label{eq:kichhoff_law}
\end{equation}
The absorption coefficient $\alpha_x$ is related to the absorption rate $\tau^{-1}$ in Eq.~\ref{eq:kompaneets_extended} as $\tau^{-1} = \alpha_x c $. 
For the source term $\epsilon (x) $ in Eq.~\ref{eq:kompaneets_extended}, we translate Eq.~\ref{eq:kichhoff_law} into photon momentum space:
\begin{equation}
 \epsilon (x) = \frac{\tau^{-1} (x) }{e^{x/\theta_e} -1}\, .
\end{equation}

\subsubsection*{Bremsstrahlung}
Thermal bremsstrahlung yields free-free (FF) absorption and emission: 
\begin{equation}
\tau_\mrm{FF}^{-1} (x)= \frac{\alpha_\mrm{f} g_\mrm{FF}(x, \theta_e)}{(24 \pi^3)^{1/2}} \lambda_C^3 \sigma_\mrm{T} c n_\mrm{e}^2 
\theta_\mrm{e}^{-\frac{1}{2}} x^{-3} (1-e^{-x/\theta_\mrm{e}}) \, ,
\end{equation}
where $\lambda_C$ is the Compton wavelength, $\alpha_\mrm{f}$ the fine structure constant, and we assume an electron-proton plasma with equal densities of protons and electrons. 
The Gaunt factor is approximated as 
\begin{equation}
g_\mrm{FF}(x, \theta_e) = \frac{\sqrt{3}}{\pi} \mrm{ln} \left( \frac{2.25 \theta_e}{x} \right)\, .
\label{eq:kirchoff_phasespace}
\end{equation}

\subsubsection*{Thermal Cyclotron}
For a cylotron energy $E_B = h e B/ (2 \pi m_e c)$, the thermal cyclotron process is characterized by
\begin{equation}
\tau_\mrm{CY}^{-1} = a \alpha_\mrm{f}^{-1} \sigma_\mrm{T} c n_\mrm{e}  \left(\frac{E_B}{(m_e c^2)}\right)^{(s-1)} \theta_e^q x^{-s} \, . 
\end{equation}
As in \citet{Vurm:2012be}, we assume
$a =10^{10}$, $q = 4$, and $s = 10$. 

\subsubsection*{Double Compton}
The double Compton process is given by
\begin{equation}
\tau_\mrm{DC}^{-1} = \frac{2 \alpha_\mrm{f}}{\pi^2 } x^{-2} \theta_e  g_\mrm{DC}(\theta_e) \lambda_C^3 \sigma_\mrm{T} c n_\mrm{e} n_\mrm{\gamma} \, ,
\end{equation}
where the Gaunt factor is 
\begin{equation}
g_\mrm{DC}(\theta_e) = (1 + 13.91 \theta_e + 11.05 \theta_e^2 + 19.92 \theta_e^3)^{-1} \, .
\end{equation}

\subsubsection*{Plasma expansion}
The  particle number conservation implies
\begin{equation}
\frac{\mrm{d}}{\mrm{d}t}(n V) = 0 \, ,
\end{equation}
which is equivalent to
\begin{align}
\frac{\mrm{d}n}{\mrm{d}t} = - \frac{n}{V}\frac{\mrm{d}V}{\mrm{d}t} = - \frac{1}{V}\frac{\mrm{d}V}{\mrm{d}\tilde{r}}\frac{\mrm{d}\tilde{r}}{\mrm{d}t} n \, .
\end{align}
We assume that  the plasma volume evolves as a power-law $V(\tilde{r}) \propto \tilde{r}^s_V$. Inserting the latter in the above equation we get
\begin{align}
\frac{\mrm{d}n}{\mrm{d}t} = - \frac{s_V}{\tilde{r}}\frac{\mrm{d}\tilde{r}}{\mrm{d}t} n \, .
\end{align}
Here, $t$ is in the plasma comoving frame and $\mrm{d} \tilde{r} / \mrm{d}t = \beta \Gamma c $. Thus, finally, 
\begin{align}
\frac{\mrm{d}n}{\mrm{d}t} = - \frac{s_V \beta \Gamma c  }{\tilde{r}}n  = - \tau^{-1}_\mrm{exp} n \, .
\end{align}

Plasma expansion also causes cooling of the plasma particles with $E \cdot V^{4/3 -1 } = const. $ for relativistic particles of energy $E$. This yields 
\begin{align}
a_\mrm{exp} (x) = \frac{\tau^{-1}_\mrm{exp}}{3} x \, .
\end{align}
Note that, in our numerical treatment, we transform to the photon momentum space, which implies a factor of $x^2$--see, \textit{e.g.}~\cite{Vurm:2008ue}.
In the following we  assume that the volume increases $\propto r^{2}$, as expected for a jet with constant opening angle. 
Note that the mass density may not evolve as $\rho \propto r^{-2}$, which we however interpret as a change of the number of particles that may for example be due to mixing with the ambient material.
As a  consequence, the expansion timescale is given by
\begin{equation}
    \tau_\mathrm{exp}^{-1} = \frac{2 \beta \Gamma c}{\tilde{r}} \, .
\end{equation}

\subsection{Numerical Treatment}
At each radius, we first evolve  the thermal electron population, then the photons and finally all other non-thermal particles. Below, we describe the numerical treatment in detail.
Throughout the simulation we choose a timestep of $0.1 t_{c} = 1/ 10 \sigma_\mrm{T} n_\mrm{e} c$.

\subsubsection{Photons}
The evolution of the photon spectra is computed with the newly developed code, \textsc{THECOS} (publically available at \url{https://zenodo.org/records/14840353} and \url{https://github.com/arudol/thecos}). 
It includes the Kompaneets equation with the additional source and sink terms, while applying a Chang \& Cooper scheme \citep{1970:ChangCooper}. To linearize the $f^2$ cooling term, we follow the description of \citet{1970:ChangCooper}, making use of the fully implicit nature of the solver; to calculate the $(k+1)$-th timestep, the cooling term is calculated from the distribution at the $k$-th timestep. The $f^2 + f$ term in Eq.~\ref{eq:kompaneets} is then given by $(f^k +1 ) f^{k+1}$. 
The equation is solved on a non-linear grid that is equally spaced in ln-space where the $i$-th grid point is defined as $x_i = \mrm{exp}((X_I +i)D_X)$. Here the constant parameter $X_I$ sets the grid origin, while $D_X$ specifies the spacing between grid points in natural logarithmic space. In our case $D_X = 0.1$, $X_I = 280$ and we adopt $580$  energy bins.
 
We start when the photon distribution is still in equilibrium with the electrons. The occupation number is given by a Planck spectrum, \ie
\begin{equation}
f(x) = (e^{x/\theta_e} -1)^{-1} \, .
\end{equation}
The energy density of a Planck spectrum is given by $u_\mrm{\gamma} = a T^4$ (with $ a = 7.56 \cdot 10^{-15}$erg/cm$^3$K$^4$), whereas the energy density of electrons is $u_\mrm{e} = \frac{3}{2} n_\mrm{e} k_\mrm{B}T$, with the $n_\mrm{e}$ being their number density. 
For a neutral thermal plasma, with an  electron-to-baryon ratio $Y_e$, the total energy density is thus $u = u_\mrm{\gamma} + u_\mrm{e} + u_\mrm{p} =  u_\mrm{\gamma} + (1+1/Y_e) u_\mrm{e}$.
In the following, we  set $Y_e = 0.5 $ as appropriate for binary neutron star merger jets \citep[see e.g.][]{Just:2022flt}.

\subsubsection{Properties of the thermal electron population}
The evolution of the plasma density $\rho$ and plasma internal energy $u$ as  functions of radius/time is given by the shell evolution extrapolated from the BNS merger simulation.
A separate PDE for the thermal electrons is not needed as long as they are in thermal equilibrium--which is the case for all scenarios investigated in this work. Hence it suffices to update at each timestep \textit{1.}~the electron number density as $n_\mrm{e} = Y_e \rho/m_\mrm{p}$ and \textit{2.}~the plasma (electron) temperature. 

The coupling between electrons and photons makes the exact calculation of the evolution of the thermal temperature  non-trivial. We instead follow a simplified approach and parametrize the cooling as $\mrm{d}\theta_e/\mrm{d}t = A ( B)$. The constants $A$ and $B$ are obtained by fitting to the thermal evolution over a wide range of radii (hence requiring that our parametrization yields the correct results for the thermal case). This approach is valid as long as the photon energy density is dominated by thermal photons and the evolution is still in the optically thick regime; both assumptions hold in our calculations.

Coulomb collisions are responsible for electron and proton relaxation to a Maxwellian distribution and for maintaining the equilibrium between electrons and protons. To calculate the Coulomb relaxation timescale we follow \cite{2007:Fundamenski}. 
They define the thermalization rate $\nu_\mathrm{Coul}$ as 
\begin{align}
\nu_\mrm{Coul (ee)} &= \frac{n_e e^4 \mrm{ln}(\Lambda_\mrm{ee})}{12 \pi^{3/2} \epsilon_\mrm{0}^2 m_e^{1/2}(k_\mrm{B}T_\mrm{e})^{3/2}}\, ,
\label{eq:nu_Coulomb_ee}
\\
\nu_\mrm{Coul (pp)} &=  \frac{n_e e^4 \mrm{ln}(\Lambda_\mrm{pp})}{12 \pi^{3/2} \epsilon_\mrm{0}^2 m_p^{1/2}(k_\mrm{B}T_\mrm{p})^{3/2}}\, , \\
\nu_\mrm{Coul (ep)} &= \frac{2^{1/2} n_e e^4 \mrm{ln}(\Lambda_\mrm{ep})}{6  \pi^{3/2} \epsilon_\mrm{0}^2 m_p^{1/2}(k_\mrm{B}T_\mrm{e})^{3/2}} \, ,
\label{eq:nu_Coulomb_ep}
\end{align}
for electron-electron (ee), proton-proton (pp) and electron-proton (ep) collisions. The respective Coulomb logarithms $\Lambda_\mrm{ee}$, $\Lambda_\mrm{pp}$ and $\Lambda_\mrm{ep}$ are  defined in \cite{2007:Fundamenski} and typically lie in the range $10$--$20$. To assess whether electrons and protons can be assumed to be in equilibrium (\textit{i.e.}~they relax into Maxwellian distribution faster than the dynamical timescale), we  compare $t_\mathrm{dyn} \equiv \tau_\mrm{exp}$ and $t_\mathrm{Coul}= 1/ \nu_\mathrm{Coul}$ (defined through Eqs.~\ref{eq:nu_Coulomb_ee}--\ref{eq:nu_Coulomb_ep}) for the typical plasma evolution. In our analytical estimate the plasma temperature $T$ was obtained through $u = u_e + u_p + u_n + u_\gamma = T [3(1 + 1/Y_e) k_\mrm{B} \rho/ 2 m_\mrm{p} + a T^3]$. This is assuming a black body distribution and equilibrium between the particle species; strictly speaking this assumption does not hold at radii larger than $10^{11}$~cm. Nonetheless it may still be indicative for the order of magnitude temperatures of the plasma and a qualitative discussion. Overall we find that equilibration can be expected for the radii considered in our calculations, namely up to  $\simeq 10^{13}$~cm and hence Maxwellian distributions can safely be assumed.

\subsubsection{Non-thermal particles and code coupling}
For the calculation of non-thermal processes we use the time-dependent sofware AM$^3$ \citep{Gao:2016uld, Klinger:2023zzv} that evolves the (coupled) evolution of neutrons, protons, electrons, positrons, neutrinos, pions, and muons. The code includes all relevant leptonic, lepto-hadronic and hadronic processes: synchrotron and inverse Compton emission  and cooling for all charged particle species, photo-pion production, photo-pair production and adiabatic cooling and density dilution due to the plasma expansion.

The evolution of the photon field is carried out by THECOS (for a detailed description see above), where the interplay between the two codes is through source/sink terms (\textsc{AM$^3$} $\rightarrow$ \textsc{THECOS}) and external fields (\textsc{THECOS} $\rightarrow$ \textsc{AM$^3$}). 
As an addition to \textsc{AM$^3$}, as described in \cite{Gao:2016uld}, several processes are updated:
\begin{enumerate}
\item Pions and muons are included as separate species, undergoing adiabatic and synchrotron losses before decaying. The treatment is similar as in \citet{Rudolph:2022ppp}.
\item Due to the large magnetic fields, a quantum synchrotron treatment is neccessary. While in \citet{Rudolph:2022ppp} this was implemented following \citet{Brainerd:1987}, we instead use a simplified approach with a cutoff of synchrotron emssion the electron energy, as suggested in \citet{Imamura:1985}. In comparison to the full treatment, this significantly increases computational speed while yielding very similar results.
\item As in \citet{Rudolph:2022ppp}, (Bethe-Heitler) photo-pair production is calculated following \citet{Kelner:2008ke}.
\item Volume expansion causes a decrease in number density and cooling of charged particles. This is implemented with a similar treatment as outlined above for photons. 
\item Proton-proton interactions are implemented following \citet{Kelner:2006tc} with the updated inelastic cross-section of \citet{Kafexhiu:2014cua}. At energies above $E_\mathrm{p}=100$~GeV, the pion distributions are calculated from the distribution functions fitted to SYBILL, for $E_\mathrm{p}<100$~GeV we use the $\delta$-function approach. The produced pions and muons below 1~GeV are decaying as described in \citet{Lipari:2007su} (without accounting for cooling effects), above 1 GeV muon and pion cooling effects are accounted for.
To obtain a smooth transition at $E_\mathrm{p}=100$~GeV (where we switch from the $\delta$-function approach to the full treatment) 
the proton inelasticity at low proton energies is set to a constant value corresponding to the one obtained from the pion redistribution function at $E_\mathrm{p}=100$~GeV.
\end{enumerate}

Note that we do not include an escape term for any particle population. This implies that the results shown for neutrinos (which can, in contrast to other particles, be assumed to free-stream out of the radiation zones) correspond to the accumulated spectra. This approach is of limited validity as we approach the photosphere and the observed spectra would instead rather correspond to the time-integrated escaped ones. Overall, this is expected to contribute to a softening of the observed spectra with respect to the results shown in this paper. 

\section{Jet conditions}
\label{appendix:jet}
Beyond the photon spectra, cooling and emission rates as well as the evolution of leptonic and hadronic particle density distributions are crucial in order to understand the main processes leading to (emitted) secondary particle spectra. This appendix shows snapshots of the particle density distributions and cooling rates for different scenarios and radii. Going from the  simplest to the most complex, we  invoke a thermal,  leptonic and finally a lepto-hadronic scenarios. 

\subsection{Thermal scenario: Jet-cocoon interface}
\begin{figure*}[hbt]
    \centering
    \includegraphics[width = \textwidth]{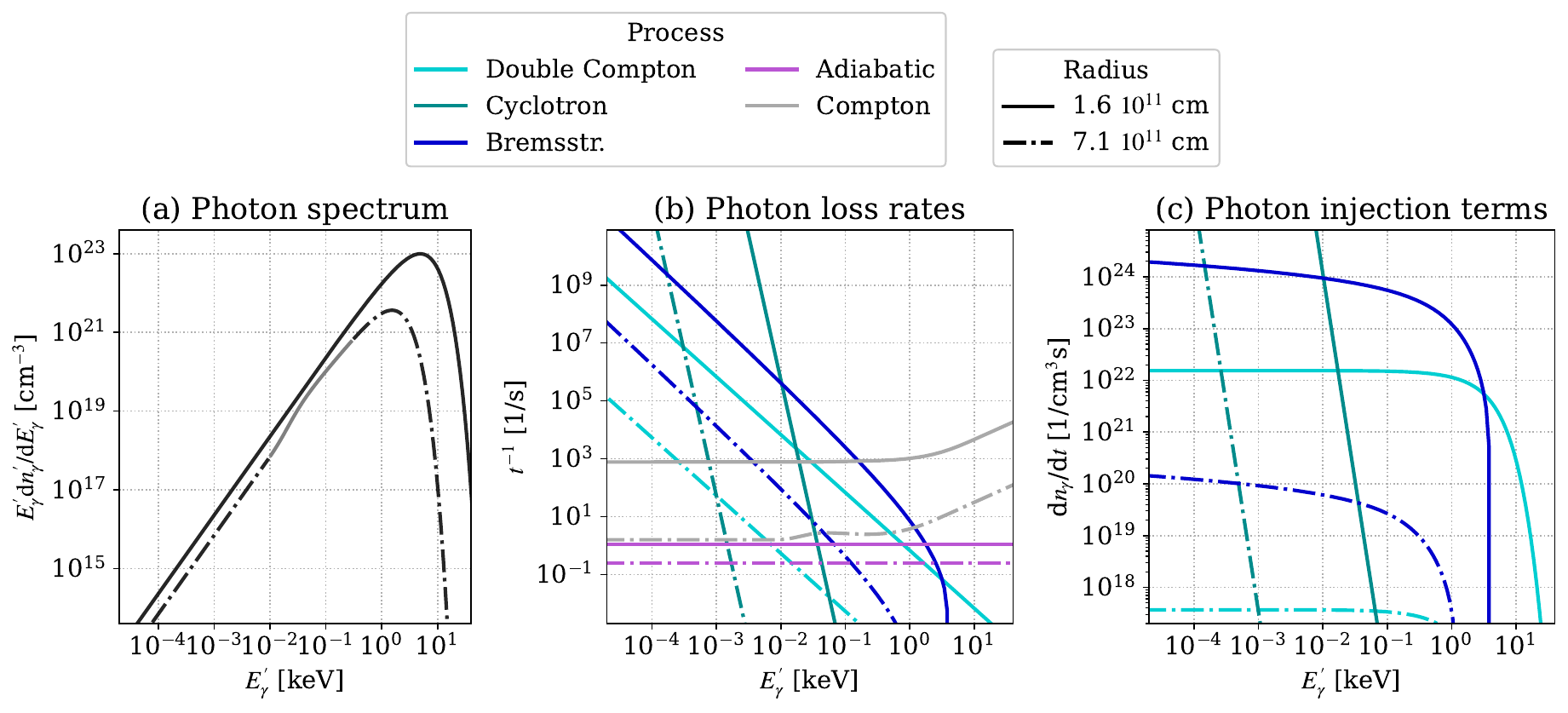} 
    \caption{Snapshots of (a) the photon spectra as well as (b) and (c) the loss/injection terms of different physical processes  at two different radii (represented by the solid and dashed-dotted  line styles). In the photon spectra plot, the grey solid line highlights the region where a slight deviation from a planck towards a Wien spectrum can be observed for the snapshot taken at $r = 7.1 \times 10^{11}$~cm.
    As an example of  thermal evolution, without additional energy dissipation, we pick the jet-cocoon interface. The Compton loss rates due to scattering on the thermal electron distribution (gray lines in the middle plot) are computed from the Kompaneets kernel. 
    }
\label{fig:thermal_photon_rates_1}
\end{figure*}

We begin with the  simplest scenario:  thermal evolution of  plasma. This is realized in the jet-cocoon interface, which we  use to illustrate the evolution from a thermal Planck to a Wien spectrum. In Fig.~\ref{fig:thermal_photon_rates_1} we show the photon spectrum as well as the loss and injection rates at two different radii for this scenario.
Let us first comment on the importance of 
different cooling processes depending on the photon energy. From the middle and right panels, we see that independent of the radius, the leading loss channel changes from cyclotron to bremsstrahlung to double Compton as the photon energy increases, with double Compton being relevant  at the highest energies only, close to the spectral peak.

The spectrum at  $r = 1.6 \times 10^{11}$~cm (solid line in the left plot) is a Planck distribution. At $r = 7.1 \times 10^{11}$~cm (dash-dotted line), a spectral break between $10^{-2}$--$10^{-1}$~keV hints at the system being in the Wien rather than in the Planck regime. This means that thermal processes do not operate fast enough with respect to the dynamical timescale of the system to ensure thermal equilibrium. This is supported by the adiabatic timescale becoming comparable to the thermal cooling  in the energy range of the spectral break. 
We point out that, in this case, the Wien spectrum is not characterized by  photon starvation (not enough photons are being produced to sustain a Planck spectrum), but by over-abundant photons.

\subsection{Leptonic scenario with magnetic reconnection: Jet with extended reconnection}

\begin{figure}[hbt]
    \centering
    \includegraphics[width = \textwidth]{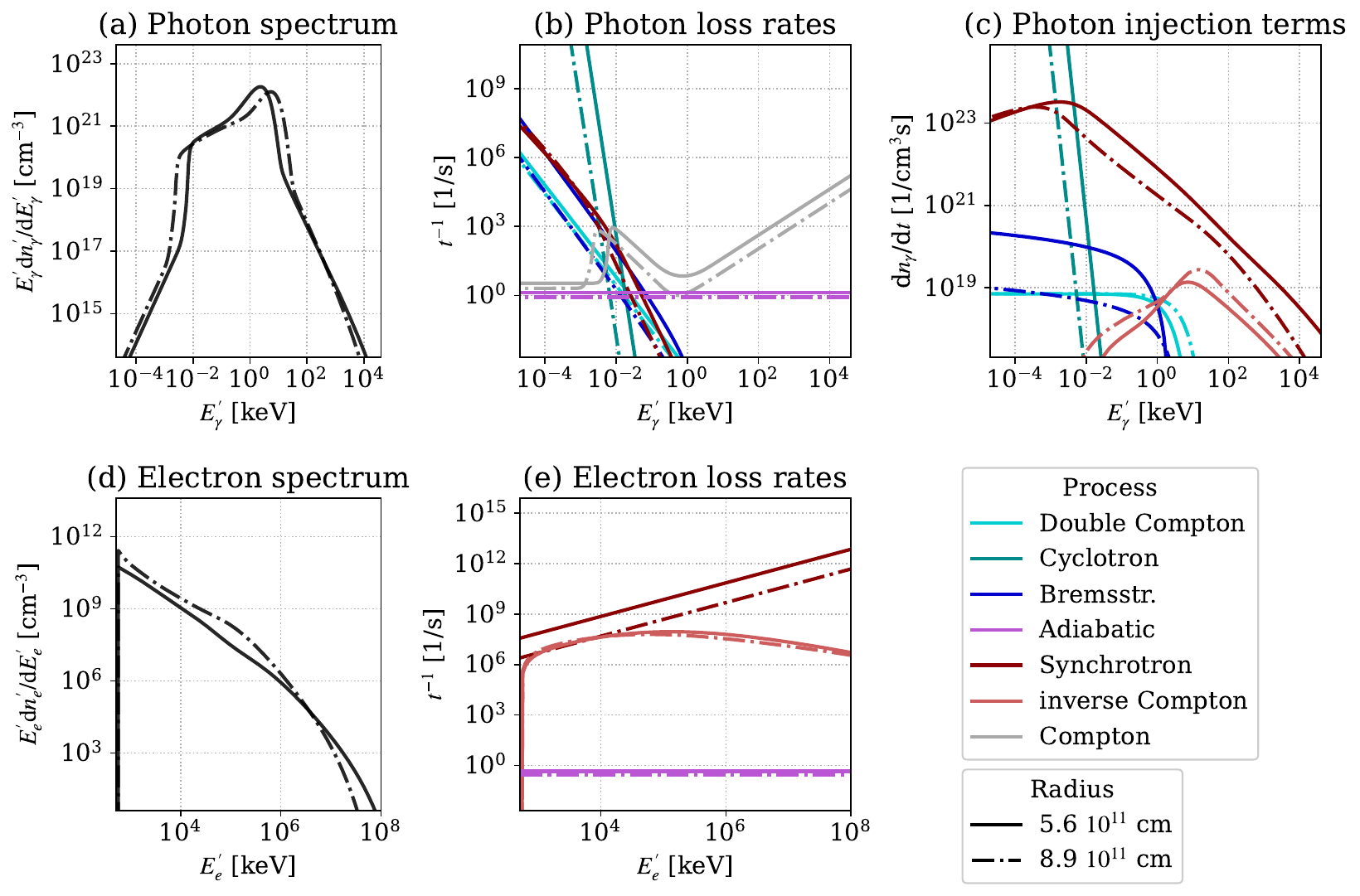} 
    \caption{Same as Fig.~\ref{fig:thermal_photon_rates_1}, including also relevant electron quantities: snapshots of (a) the photon spectra, (b) photon loss rates, (c) photon injection terms, (d) electron spectra and (e) electron loss rates. As an example of  evolution, where magnetic reconnection leads to injection of accelerated electrons, in addition to the thermal distribution, we consider the jet with extended reconnection. Both snapshots are taken at radii where  reconnection is still active. The Compton loss rates, due to scattering on the thermal electron distribution (gray lines in the middle plot), are computed from the Kompaneets kernel, while inverse Compton scatterings (light red lines) affect the non-thermal electron distribution. Synchrotron emission cools the electrons, while photon losses includes synchrotron-self absorption on non-thermal electrons. }
\label{fig:thermal_photon_rates_2}
\end{figure}
Adding accelerated electrons, the photon spectrum at the lowest energies is in thermal equilibrium, as shown in Fig.~\ref{fig:thermal_photon_rates_2}. Here cyclotron losses are   the most efficient emission and absorption process. Then, around $10^{-2}$~keV, synchrotron losses kick in and are dominant up to the highest energies (although with a small contribution from inverse Compton emission). 
Due to efficient Comptonization, we expect the spectrum to be 
reshaped both below and above the spectral peak.
Note that synchrotron self-absorption losses are also relevant.

Accelerated electrons  are injected at both selected  radii, with  the maximum energy being slightly lower at the larger radius, with  low-energy electrons  piling up.
The most efficient cooling process is the synchrotron one at high energies; at low energies inverse Compton scatterings in the Thomson regime is as  efficient as  synchrotron
at the  largest selected  radius. Both synchrotron and inverse Compton are well above the adiabatic cooling rate, hence electrons are in the fast-cooling regime.

\subsection{Lepto-hadronic scenario with sub-shocks: Outer Jet}
\begin{figure}[hbt]
    \centering
    \includegraphics[width = 0.8\textwidth]{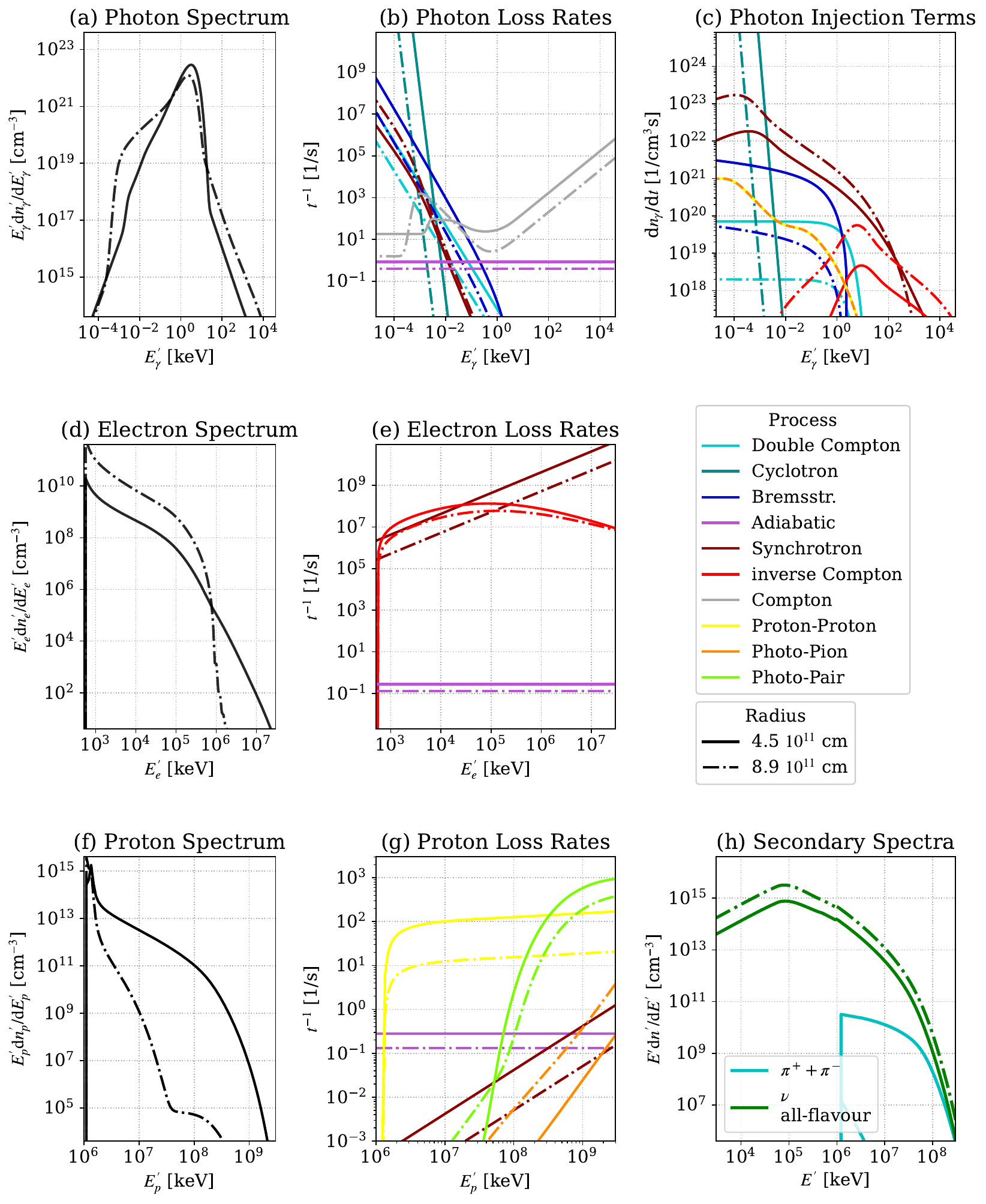} 
    \caption{Same as Fig.~\ref{fig:thermal_photon_rates_1}, including also relevant electron and hadronic quantities: snapshots of (a) the photon spectra, (b) photon loss rates, (c) photon injection terms, (d) electron spectra, (e) electron loss rates, (f) proton spectra, (g) proton loss rates and (h) secondary spectra. As an example of  evolution where collisionless subshocks lead to injection of accelerated electrons and protons in addition to the thermal distributions we take the outer jet scenario. The first (second) snapshot is taken whithin (slightly above) the subshock region. Note that Compton loss rates due to scattering on the thermal electron distribution (gray lines in the middle plot) are computed from the Kompaneets kernel, while inverse Compton scatterings (dark red lines) affect the non-thermal electron distribution. In the photon injection plot, the dashed yellow-orange curve corresponds to the summed contribution of all secondary particles stemming from  pions (produced in lepto-hadronic and hadronic interactions) at $4.5 \times 10^{11}$~cm. At the second snapshot, $8.9 \times 10^{11}$~cm, these processes have negligible contribution and hence are not shown. 
    }
\label{fig:thermal_photon_rates_3}
\end{figure}
The impact of both leptonic and hadronic processes 
can be illustrated relying on the the outer jet where collisionless sub-shocks co-accelerate both particle species (cf.~Fig.~\ref{fig:thermal_photon_rates_3}). The first snapshot at $4.5 \times 10^{11}$~cm is within the subshock region (where particle acceleration is active), the second one at $8.9 \times 10^{11}$~cm above the subshock region (where thus particle acceleration has already ceased). 

At both radii, the photon spectrum is significantly altered with respect to the  thermal one: Comptonization of additional seed photons broadens the spectrum below the thermal peak, while an additional power-law appears above the peak. Overall this spectral shape is similar to the purely leptonic spectrum. This may be due  to the fact that similar processes  take place--photon losses are dominated by thermal losses at lower energies and adiabatic losses at the higher ones; in addition to efficient Compton scatterings. Photon injection is predominantly determined by  cyclotron emission at the lowest energies, while non-thermal electron synchrotron and inverse Compton scatterings take over with increasing photon energy. Again, the inverse Compton component takes place at energies that are too high to be relevant. 

Electrons efficiently cool via inverse Compton/synchrotoron emission. The lower magnetic field in the outer jet compared to the on-axis case increases the significance of inverse Compton cooling with respect to the synchrotron cooling. 
Interestingly, although acceleration has ceased at the selected larger  radius, a relevant non-thermal electron population is still present. 

The proton distribution is significantly diminished at the larger radius. Nonetheless, hadronic interactions may still produce a small number of electrons even at the larger radius.
Protons cool efficiently through proton-proton interactions at lower energies, with Bethe-Heitler pair production taking place at higher energies. 
As for  the secondaries of hadronic interactions, due to the fact that we neglect particle escape from the production region in our model, neutrinos ``pile up'' due to hadronic interactions. Thus, their spectra  are similar to time-integrated spectra obtained within  a realistic scenario with  free-streaming neutrinos. 
Pions are mostly present at the smaller radius where energetic protons are abundantly avalaible for their production through hadronic interactions.

\bibliography{references}{}
\bibliographystyle{aasjournal}

\end{document}